\UseRawInputEncoding
\documentclass[fleqn,usenatbib]{mnras}
\usepackage[T1]{fontenc}
\DeclareRobustCommand{\VAN}[3]{#2}
\let\VANthebibliography\thebibliography
\def\thebibliography{\DeclareRobustCommand{\VAN}[3]{##3}\VANthebibliography}
\usepackage{graphicx}	% Including figure files
\usepackage{amsmath}	% Advanced maths commands
\usepackage{amssymb}	% Extra maths symbols
\usepackage{multirow}
\usepackage{threeparttable}
\usepackage{tablefootnote}
\title[Theoretical uncertainties for weak lensing]{
Assessing theoretical uncertainties for cosmological constraints from weak lensing surveys}

\author[Tan T. et al.]{
Ting Tan,$^{1,2}$\thanks{E-mail: ting.tan@lpnhe.in2p3.fr}
Dominik Z\"urcher,$^{2}$
Janis Fluri,$^{2,3}$ Alexandre Refregier,$^{2}$
Federica Tarsitano,$^{2}$
\newauthor{and Tomasz Kacprzak$^{2}$}
\\
$^{1}$Sorbonne Universit\'e, CNRS/IN2P3, Laboratoire de Physique Nucl\'eaire et de Hautes Energies, LPNHE, 
\\4 Place Jussieu, F-75252 Paris, France\\
$^{2}$Institute for Particle Physics and Astrophysics, Department of Physics, ETH Z\"urich,
\\Wolfgang Pauli Strasse 27, 8093 Z\"urich, Switzerland\\
$^{3}$Data Analytics Lab, Department of Computer Science, ETH Zurich\\ Universit\"atstrasse 6, 8006 Z\"urich, Switzerland
}

\date{Accepted XXX. Received YYY; in original form ZZZ}

\pubyear{2021}

\begin{document}
\label{firstpage}
\pagerange{\pageref{firstpage}--\pageref{lastpage}}
\maketitle

\begin{abstract}
Weak gravitational lensing is a powerful probe which is used to constrain the standard cosmological model and its extensions. With the enhanced statistical precision of current and upcoming surveys, high accuracy predictions for weak lensing statistics are needed to limit the impact of theoretical uncertainties on cosmological parameter constraints.  
%Weak gravitational lensing is a powerful cosmological probe . The weak lensing angular power spectrum can indeed be used to constrain and test the cosmological model from weak lensing surveys. Since it is predicted from the matter power spectrum, the accurate theoretical prediction of the matter power spectrum is essential for such studies. 
For this purpose, we present a comparison of the theoretical predictions for the nonlinear matter and weak lensing power spectra, based on the widely used fitting functions ($\texttt{mead}$ and $\texttt{rev-halofit}$), emulators ($\texttt{EuclidEmulator}$, $\texttt{EuclidEmulator2}$, $\texttt{BaccoEmulator}$ and $\texttt{CosmicEmulator}$) and N-body simulations ($\texttt{Pkdgrav3}$). We consider the forecasted constraints on the $\Lambda \texttt{CDM}$ and $\texttt{wCDM}$ models from weak lensing for stage \uppercase\expandafter{\romannumeral3} and  stage \uppercase\expandafter{\romannumeral4} surveys. We study the relative bias on the constraints and their dependence on the assumed prescriptions. Assuming a $\Lambda \texttt{CDM}$ cosmology, we find that the relative agreement on the $S_8$ parameter is between $0.2-0.3\sigma$ for a stage \uppercase\expandafter{\romannumeral3}-like survey between the above predictors. For a stage \uppercase\expandafter{\romannumeral4}-like survey the agreement becomes $1.4-3.0\sigma$. 
In the $\texttt{wCDM}$ scenario, we find broader $S_8$ constraints, and agreements of $0.18-0.26\sigma$ and $0.7-1.7\sigma$ for stage \uppercase\expandafter{\romannumeral3} and stage \uppercase\expandafter{\romannumeral4} surveys, respectively.
The accuracies of the above predictors therefore appear adequate for stage \uppercase\expandafter{\romannumeral3} surveys, while the fitting functions would need improvements for future stage \uppercase\expandafter{\romannumeral4} weak lensing surveys. Furthermore, we find that, of the fitting functions, $\texttt{mead}$ provides the best agreement with the emulators. We discuss the implication of these findings for the preparation of the future weak lensing surveys.
\end{abstract}

% Select between one and six entries from the list of approved keywords.
% Don't make up new ones.

%%%%%%%%%%%%%%%%%%%%%%%%%%%%%%%%%%%%%%%%%%%%%%%%%%

%%%%%%%%%%%%%%%%% BODY OF PAPER %%%%%%%%%%%%%%%%%%

\section{Introduction}

The next generation of wide field cosmological surveys, such as LSST \footnote{\url{https://www.lsst.org}}, Euclid \footnote{\url{https://www.cosmos.esa.int/web/euclid/home}}, and NGRST \footnote{\url{https://roman.gsfc.nasa.gov/}} will map the matter distribution of the local Universe with an unprecedented accuracy. These high precision measurements present a challenge for the theoretical modeling of cosmological observables. Cosmic shear is a cosmological observable 
that relies on the distortions of galaxy shapes caused by weak gravitational lensing \citep[eg.][]{bartelmann2001weak}. This effect is due to the gravitational deflection of photons by the matter density field along the line of sight. Cosmic shear measures the inhomogeneities in the cosmic density field with high precision and can be used as an unbiased tracer of the matter distribution. It is sensitive to both, the matter distribution of the Universe and the growth of cosmic structure, which is important for the understanding of the expansion history of the Universe. A commonly used cosmic shear summary statistic is the cosmic shear angular power spectrum, which can be predicted from the matter power spectrum. The modeling of the matter power spectrum on large scales can be derived using perturbation theory \citep{PTBlas_2014,PTBERNARDEAU20021,PTPhysRevD.73.063519,PThttps://doi.org/10.1111/j.1365-2966.2012.22127.x,PTBlas_2016,PTNISHIMICHI2016247,PTBaumann_2012,PTcataneo2019road,PTforeman2016eft,PTbeutler2017clustering,PTheory2020}, where the structure formation of the Universe is linear.
However, at non-linear, small scales with $k\gtrsim 1h\text{Mpc}^{-1}$, nonlinear processes have a strong impact on the matter power spectrum, and perturbation theory is no longer valid. 

\noindent In this work, we compare the theoretical predictions of the nonlinear matter power spectrum, and the associated theoretical uncertainties on cosmological parameters from measurements of the cosmic shear angular power spectrum. The comparison includes some widely used models fitted from N-body simulations using analytical halo models: $\texttt{halofit}$ \citep{10.1046/j.1365-8711.2003.06503.x} is fitted to low resolution, gravity-only N-body simulations, which is known to exhibit a non-negligible mismatch with current state-of-the-art hydrodynamic N-body simulations; $\texttt{rev-halofit}$ \citep{Takahashi_2012}, developed as the revisited version of $\texttt{halofit}$ is used in the analysis of the Dark Energy Survey (DES) \citep{PhysRevD.98.043528}; and $\texttt{mead}$ \citep{10.1093/mnras/stv2036}, which is used in the analysis of the Kilo-Degree Survey (KiDS) combined with the VISTA Kilo-Degree Infrared Galaxy Survey (VIKING) \citep{hildebrandt2020kids+}. Apart from the halo model fitting method, emulators are generated from the interpolation of a suite of N-body simulations, e.g. $\texttt{CosmicEmulator}$ \citep{heitmann2009coyote,lawrence2017mira,heitmann2013coyote}, $\texttt{BaccoEmulator}$ \citep{angulo2020bacco,arico2021bacco}, $\texttt{EuclidEmulator}$ \citep{Euclid2019} and its updated version $\texttt{EuclidEmulator2}$ \citep{euclidcollaboration2020euclid}, $\texttt{COSMOPOWER}$ \citep{mancini2021cosmopower} and $\texttt{GP emulator}$ \citep{giblin2019road}. In this study, $\texttt{CosmicEmulator}$, $\texttt{BaccoEmulator}$, $\texttt{EuclidEmulator}$ and $\texttt{EuclidEmulator2}$ are representatively selected in the comparison at the level of the matter power spetrum, and a comparison between $\texttt{rev-halofit}$, $\texttt{mead}$ and $\texttt{EuclidEmulator}$ is also shown in \citet{knabenhans2021parameter}. In order to estimate the theoretical uncertainties, we look at the weak lensing cosmological parameter constraints, by generating a forecast for a stage \uppercase\expandafter{\romannumeral3}, DES-like survey and a stage \uppercase\expandafter{\romannumeral4}, Euclid-like survey. We take into account the parameters described by the standard $\Lambda \text{CDM}$ cosmological model and the extended $w\text{CDM}$ model.
\newline
This paper is organised as follows. In Section \ref{sec2} we describe the theoretical framework, including three halo-model based fitting functions, $\texttt{mead}$, $\texttt{halofit}$ and $\texttt{rev-halofit}$; four power spectrum emulators extracted from N-body simulations: $\texttt{CosmicEmulator}$, $\texttt{BaccoEmulator}$, $\texttt{EuclidEmulator}$ and $\texttt{EuclidEmulator2}$, and one N-body simulation code $\texttt{Pkdgrav3}$ \citep{potter2017pkdgrav3}. In Section \ref{sec:3} we present the method and the relevant codes used in this study. We summarize our results in Section \ref{sec4} and our conclusions in Section \ref{sec5}.

\section{Theory}\label{sec2}
In this section, we describe the theoretical background of the matter power spectrum, weak lensing and its angular power spectrum, as well as the different predictors of the matter power spectrum that we include in the comparisons.
\subsection{Weak Lensing}\label{sec2.1}
Considering the cosmic density field $\rho(\Vec{r})$ at the position $\Vec{r}$, the density contrast $\delta(\Vec{r})$ is defined as the relative difference of $\rho(\Vec{r})$ to the average density $\Bar{\rho}$
\begin{equation}
    \delta(\Vec{r})=\frac{\rho(\Vec{r})-\Bar{\rho}}{\Bar{\rho}}.
\end{equation}
In Fourier space, the density contrast takes the following form
\begin{equation}
    \delta(\Vec{k})=\int \delta(\Vec{r})\exp{(i\Vec{k}\cdot\Vec{r})}\rm{d}^3r.
\end{equation}
Furthermore, the matter power spectrum $P(\Vec{k})$ is defined as the correlation of the density contrast in Fourier space \citep{peebles1980large}:
\begin{equation}
    \langle\delta(\Vec{k})\delta(\Vec{k}')\rangle=(2\pi)^3\delta_{\rm{D}}^{(3)}(\Vec{k}+\Vec{k}')P(\Vec{k}),
\end{equation}
where $\delta_{\rm{D}}$ is the three dimensional Dirac delta function.\newline\newline
For full-sky surveys, the cosmic shear angular power spectrum is approximately identical to the convergence power spectrum \citep{bartelmann2016weak}, which can be defined as a weighted integration along the line-of-sight over the matter power spectrum \citep{bartelmann2001weak}, and simplified using the Kaiser-Limber approximation \citep{PhysRevD.78.123506,1953ApJ...117..134L,1992ApJ...388..272K,1998ApJ...498...26K}. We follow the formalism of \citep{PhysRevD.78.123506,kitching2017limits,kilbinger2017precision,tarsitano2020predicting,giannantonio2012constraining} to compute the cross-correlated shear power spectrum with tomographic redshift bins $i$ and $j$:
\begin{equation}
    C^{ij}_\gamma (\ell)=\frac{9}{16}\left(\frac{H_0}{c}\right)^4 \Omega^2_{\rm{m}} \int_0^{\chi_{\rm{h}}} \rm{d}\chi P_{\text{NL}}\left(\frac{\ell}{r},\chi\right)\frac{g_i(\chi)g_j(\chi)}{(ar(\chi))^2}
\label{Equ:shear Cls}
\end{equation}
Here $P_{\text{NL}}$ is the non linear matter power spectrum, $\chi$ is the comoving distance, $\chi_{\rm{h}}$ is the comoving horizon distance, $\Omega_{\rm{m}}$ is the total matter density,  $a=(1+z)^{(-1)}$ is the scale factor and $g(\chi)$ is the lensing efficiency function defined as:
\begin{equation}
    g_i(\chi) = 2\int^{\chi_{\rm{h}}}_\chi \rm{d}\chi' n_i(\chi)\frac{r(\chi)r(\chi'-\chi)}{r(\chi')},
\end{equation}
with $n_i(\chi)$ being the normalized number density of the observed galaxies at a comoving distance $\chi$.
\subsection{Matter Power Spectrum}
The matter power spectrum is a fundamental statistics to study the large scale structure of the Universe. As seen above, it is, in particular, useful to predict the cosmic shear angular power spectrum. Therefore, it is necessary to have an accurate theoretical model for the matter power spectrum on all scales. On large scales and mildly non-linear scales, the matter power spectrum can be modeled using perturbation theory and some extended theories \citep{2021Euclid}. On small scales, which are in the non-linear regime, these approaches are not suited to predict the power spectrum with the necessary precision, while other methods are developed with the use of halo model or simulations.
\subsubsection{Analytical Predictions}\label{halofit codes}
A common way to model the matter power spectrum on these small scales is to empirical fit physically motivated formulas to measurements from N-body simulations, e.g. as done in \citet{1991ApJ...374L...1H}. Furthermore, modelling the density field as a collection of virialized halos, the matter power spectrum can be approximated analytically using the statistics of halos, and fitted to simulations or emulators \citep{2000ApJ...543..503M,10.1046/j.1365-8711.2000.03715.x,COORAY20021}. 

\noindent In this study, we compare 3 halo-model based fitting functions: $\texttt{mead}$, $\texttt{halofit}$, and $\texttt{rev-halofit}$. $\texttt{halofit}$ was built using a series of N-body simulations with a total of $N=256^3$ particles and the box size from $84\text{Mpc}/h$ to $240\text{Mpc}/h$. Using the halo model, the matter power spectrum is constructed with two terms, the one-halo term proposed by \citet{10.1046/j.1365-8711.2000.03779.x,2000ApJ...543..503M,10.1046/j.1365-8711.2000.03715.x,Scoccimarro_2001} and a two-halo term \citep{2000ApJ...543..503M,10.1046/j.1365-8711.2000.03715.x,Scoccimarro_2001} to describe the exclusion effects between dark matter halos. The one-halo term indicates the correlation of the matter field of one single halo, which dominates on small scales, while the two-halo term describes the cross-correlation between different halos, that has a strong impact on larger scales. Assuming that the halos are distributed according to the halo mass function \citep{1974ApJ...187..425P,1999MNRAS.308..119S}, the matter power spectrum modelled with this approach can achieve a high precision on large scales. However, due to the lack of baryons and the relatively low resolution of the N-body simulations used in their study, $\texttt{halofit}$ does not match high resolution N-body simulations, giving an accuracy at the $5\%$ level at $k=1h\text{Mpc}^{-1}$ \citep{heitmann2010coyote}, and larger differences for $k>1h\text{Mpc}^{-1}$, which is insufficient for the non-linear regime. $\texttt{rev-halofit}$ is a revised prescription of $\texttt{halofit}$, which provides a more accurate prediction of the matter power spectrum for $k<30h\text{Mpc}^{-1}$ and $z<10$, with a $5\%$ level accuracy at $k=1h\text{Mpc}^{-1}$ and $10\%$ level accuracy at $k=10h\text{Mpc}^{-1}$. $\texttt{rev-halofit}$ uses high resolution N-body simulations for 16 cosmological models around the Wilkinson Microwave
Anisotropy Probe (WMAP) best-fit cosmological parameters. The N-body simulations were run with the $\texttt{Gadget-2}$ N-body code \citep{gadget,gadget-2}, $1024^3$ particles in total, and the box size from $320\text{Mpc}/h$ to $2000\text{Mpc}/h$. The power spectrum is fitted using an improved fitting formula with 5 more model parameters as compared to $\texttt{halofit}$. Several extended methods have been proposed to improve the halo model \citep{10.1111/j.1365-2966.2011.20222.x,10.1093/mnras/stu1972,PhysRevD.91.123516}. Here we only consider $\texttt{mead}$ \citep{10.1093/mnras/stv2036}, which reaches an accuracy at the $5$ percent level for $k=10h\text{Mpc}^{-1}$ and $z<2$. $\texttt{mead}$ introduces more physical parameters in addition to the halo model, and is fitted to the ``Coyote Universe''\citep{heitmann2013coyote} suite of high resolution simulations, the same simulations used for the generation of $\texttt{CosmicEmulator}$. It also includes massive neutrinos \citep{10.1093/mnras/stw681} and baryonic effects e.g. active galactic nuclei (AGN) feedback, supernovae explosions, and gas cooling. However, we only consider the dark-matter-only case in this study. 

\subsubsection{Emulators}
The fitting functions based on halo models described in Section~\ref{halofit codes} can provide accurate non-linear power spectrum predictions for large $k$-modes and a wide redshift range, which can be used to predict cosmological observables. However, they also have limitations as the precision is not uniform for different cosmological parameters, and it is difficult for fitting functions to give a high precision below the $1\%$ level compared to high resolution simulations. Power spectrum emulators are constructed following a different approach in which one interpolates the power spectrum from a set of N-body simulations within a certain range of relevant parameters, using interpolation methods, e.g. Gaussian Processes
Regression \citep{heitmann2010coyote,heitmann2013coyote,angulo2020bacco} or polynomial chaos expansion \citep{Euclid2019,euclidcollaboration2020euclid}. Compared to fitting functions, emulators usually provide consistent precision of the predictions for different $k$-modes. However, emulators also have limitations: Firstly, the covered parameter space is limited, thus making it difficult to perform a likelihood analysis, for which one needs to explore a wide range of parameter values. Secondly, the ranges of $k$ and redshift are also limited, making it difficult to compute the weak lensing cosmic shear observables for high $\ell$s, which requires an integration over a large $k$ range. 

\noindent In this study, we compare 4 emulators: $\texttt{CosmicEmulator}$ \citep{Heitmann_2016}, $\texttt{BaccoEmulator}$ \citep{angulo2020bacco}, $\texttt{EuclidEmulator}$ \citep{Euclid2019}, and $\texttt{EuclidEmulator2}$ \citep{euclidcollaboration2020euclid}. $\texttt{CosmicEmulator}$ is fitted using a set of the ``Coyote Universe" simulations and the "Mira-Titan Universe"\citep{lawrence2017mira} simulations. We use the latest version of the emulator \citep{Heitmann_2016}, for which the ``Mira-Titan Universe" simulations were run with $3200^3$ particles and a simulation volume of $(2100 h^{-1}\text{Mpc})^3$. The $\texttt{CosmicEmulator}$ successfully achieves high precision predictions of the power spectrum within the $4\%$ level for $ k_{max}=5h\text{Mpc}^{-1}$ and $z<2$. It allows for the variation of various parameters, including
the matter density $\Omega_{\rm{m}}$, the amplitude of density fluctuations $\sigma_8$, the baryon density $\Omega_{\rm{b}}$, the scalar spectral index $n_{\rm{s}}$, the dark energy equation of state parameters $w_0$ and $w_{\rm{a}}$, the dimensionless Hubble parameter $h$, the neutrino density $\Omega_\nu$, and the redshift $z$. $\texttt{EuclidEmulator}$ uses a different emulation method using N-body simulations generated with the $\textsc{PkdGrav3}$ code \citep{potter2017pkdgrav3}. It uses $100$ simulations with $2048^3$ particles in a $(1250h^{-1}\text{Mpc})^3$ simulation volume. The non-linear correction is encoded as a boost factor adding up to the input linear power spectrum, achieving a precision at the $1\%$ level for predictions within the ranges $k<1hMpc^{-1}$ and $z<1$. \citet{Euclid2019} demonstrated that $\texttt{EuclidEmulator}$ agrees with $\texttt{rev-halofit}$ at the $8\%$ level. As an updated version of $\texttt{EuclidEmulator}$, $\texttt{EuclidEmulator2}$ is extended with dynamical dark energy and massive neutrinos, created with a larger parameter space and a modified version of the $\texttt{Pkdgrav3}$ N-body code. $\texttt{EuclidEmulator2}$ provides a consistent accuracy with simulations at the $2\%$ level up to $k_{max}=10h\text{Mpc}^{-1}$ for $z<2$, and slightly lower accuracy for higher redshift $z\sim3$. However, as $\texttt{EuclidEmulator2}$ uses the amplitude of the primordial power spectrum $A_{\rm{s}}$ instead of $\sigma_8$ as input parameter, we use the following formula to transfer $\sigma_8$ into $A_{\rm{s}}$\citep{Hand_2018}:
\begin{equation}
    A_{\rm{s}}=\left(\frac{\sigma_8}{\sigma_{8,0}}\right)^2\times A_{\rm{s},0}
\end{equation}
in our comparison, where $\sigma_{8,0}=0.826$ and $A_{\rm{s},0}=2.184\times10^{-9}$.
\newline
$\texttt{BaccoEmulator}$ is another state-of-the-art emulator using an updated version of the $\texttt{L-Gadget3}$ code \citep{gedget2,LGadget3} with $4320^3$ particles in a $(1440h^{-1}\text{Mpc})^3$ simulation volume. It has a $2\%$ level accuracy over the redshift range $0<z<1.5$ and $k<5h\text{Mpc}^{-1}$.

\subsubsection{N Body Simulations}
We also include in this study a comparison with a dark-matter-only N-body simulation run with $\textsc{PkdGrav3}$, which is based on a binary tree algorithm. This code uses $5^{\rm th}$ order multipole expansions of the gravitational potential between particles and can achieve fast computational speeds with hardware acceleration. A comparison between \textsc{PkdGrav3} and the N-body codes, $\texttt{Gadget-3}$, $\texttt{Gadget-4}$ and $\texttt{Ramses}$ is presented in \citet{Schneider_2016} and \citet{springel2020simulating}. The $\textsc{PkdGrav3}$ simulations are the same as the ones used for $\texttt{EuclidEmulator}$, with $2048^3$ particles in total and the box size of $L=1250h^{-1}\text{Mpc}$. The details are presented in \citet{Euclid2019}.

\section{Method}\label{sec:3}
\begin{table}
\caption{Parameter settings for mock surveys: The Stage \uppercase\expandafter{\romannumeral4} survey is created using a 4 times larger survey area and galaxy density compared to the Stage \uppercase\expandafter{\romannumeral3} Survey. A deeper Smail redshift distribution is also used in the Stage \uppercase\expandafter{\romannumeral4} Survey.}
%\resizebox{\columnwidth}{!}
%\begin{threeparttable}
\begin{tabular}{lll}
\hline
Survey&Stage \uppercase\expandafter{\romannumeral3}&Stage \uppercase\expandafter{\romannumeral4}
\\
\hline
Survey Area [deg$^{2}$] &5'000&20'000\\
\hline
Galaxy Density [arcmin$^{-2}$] &5&20\\
\hline
Redshift Distribution&Smail& Smail\\
\hline
Redshift Bins&4&4\\
\hline
Redshift Range\tablefootnote{The presented redshift range refers to the considered range used in the generation process of the covariance matrix for the mock surveys. The range differs from the redshift range used for the predictors of the weak lensing power spectrum in Section~\ref{sec:4.2}, where we use $[0.08,2]$ for the stage \uppercase\expandafter{\romannumeral3} survey and $[0.08,3.0]$ for the stage \uppercase\expandafter{\romannumeral4} survey.}&$0.025\sim3.0$&$0.025\sim3.0$\\
%\tnote{1}
\hline
\end{tabular}
\label{tab:parameter_survey}
%\begin{tablenotes}
%\footnotesize
%\item[1] The presented redshift range refers to the considered range used in the generation process of the covariance matrix for the mock surveys. The range differs from the redshift range used for the predictors of the weak lensing power spectrum in Section~\ref{sec:4.2}, where we use $[0.08,2]$ for the stage \uppercase\expandafter{\romannumeral3} survey and $[0.08,3.0]$ for the stage \uppercase\expandafter{\romannumeral4} survey.
%\end{tablenotes}
%\end{threeparttable}
\end{table}

In this work, we perform a comparison of predictors of the nonlinear matter power spectrum, i.e. halo-model based fitting functions and emulators. We estimate the theoretical uncertainties of these predictors on the parameter constraint level by looking at the weak lensing cosmological parameter constraints from a stage \uppercase\expandafter{\romannumeral3} survey and a stage \uppercase\expandafter{\romannumeral4} survey. For each survey, we perform a comparison using the standard $\Lambda \text{CDM}$ cosmological model and the extended $w\text{CDM}$ model.
\subsection{Survey parameters}\label{sec:3.1}
The estimate of the theoretical uncertainties for cosmological parameters is realised by forecasting the constraints for a stage \uppercase\expandafter{\romannumeral3} survey and a stage \uppercase\expandafter{\romannumeral4} survey. The covariance matrix is estimated from simulations, using the \texttt{NGSF} code described in \citet{zurcher2021cosmological} and \citet{zuercher2021peaks}. We refer the reader to \citet{zurcher2021cosmological} for a detailed description of the method. Table~\ref{tab:parameter_survey} shows the parameter settings used for the generation of the mock galaxy surveys. \citet{2021Euclid} suggests using $\ell_{\rm{max}}=5000$ for stage \uppercase\expandafter{\romannumeral4}-like surveys to probe deep into non-linear regime. However, in this study we use a more conservative limit of $\ell_{\rm{max}}=1000$, and do not take into account baryonic effects.\\

\begin{figure}
\centering
\includegraphics[width=8.1cm,height=5.4cm]{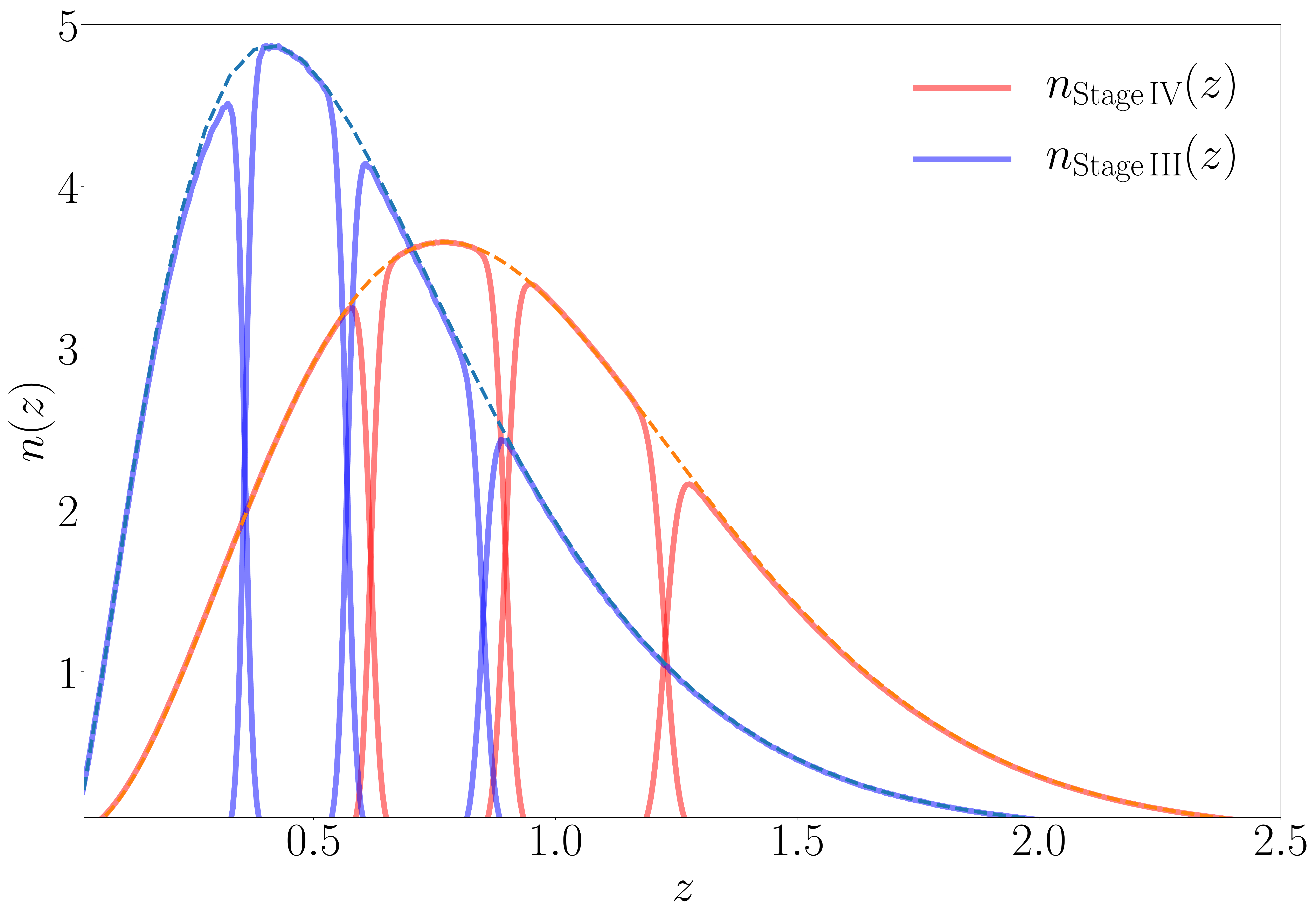}
\caption{The redshift distributions of the source galaxies. One can see the four tomographic distributions for the stage \uppercase\expandafter{\romannumeral3} survey and the stage \uppercase\expandafter{\romannumeral4} survey. The global distributions, that follows the \citet{smail1995deep} model, is shown by the dashed lines.}
\label{fig:redshift_distribution}
\end{figure}

%\subsection{Tomographic redshift distributions}\label{sec:3.1}
\noindent We use \citet{smail1995deep} distributions to model the global redshift distribution of the source galaxies for both the stage \uppercase\expandafter{\romannumeral3} survey and the stage \uppercase\expandafter{\romannumeral4} survey. The corresponding  formulas and parameter settings for these two distributions are as follows
\begin{equation}
    n(z)_{\text{stage\uppercase\expandafter{\romannumeral3}} }=z^\alpha \exp{[-(\frac{z}{z_0})^\beta]},
\end{equation}
with $\alpha=1.5$, $\beta=1.1$ and $z_0=0.31$ and
\begin{equation}
    n(z)_{\text{stage\uppercase\expandafter{\romannumeral4}}}=(\frac{z}{z_0})^\alpha \exp{[-(\frac{z}{z_0})^\beta]},
\end{equation}
with $\alpha=2.0$, $\beta=1.5$ and $z_0=0.64$ \citep{2021Euclid}. In both cases the source galaxies are divided into four tomographic bins with equal number of galaxies in each bin. As a result of the auto- and cross-combinations of these four redshift bins, we have 10 combinations of auto- and cross-correlations for the cosmic shear measurements (4 auto-correlations and 6 cross-correlations). Figure~\ref{fig:redshift_distribution} shows the global and tomographic redshift distributions used in this study.

\subsection{Covariance Matrix}\label{sec:3.2}
An accurate estimate of the survey covariance matrix is crucial for the correct calculation of the likelihood function. We estimate the covariance matrices for the stage \uppercase\expandafter{\romannumeral3} and stage \uppercase\expandafter{\romannumeral4} survey setups described in Table~\ref{tab:parameter_survey} from numerical simulations. We generate a large number ($N = 2000$) of realisations of the angular power spectra for each survey setup following the methodology outlined in \citet{zurcher2021cosmological}. In the following, we introduce the used N-Body simulations, briefly summarize the forward modelling procedure used to generate the angular power spectra and describe the estimation of the covariance matrix. We refer the reader to \citet{zurcher2021cosmological} for a more detailed description of the methodology. \\

\noindent We utilise the 50 independent \textsc{PkdGrav3} \citep{potter2017pkdgrav3} N-Body simulations at the fiducial cosmology that were previously used in \citet{zurcher2021cosmological, zuercher2021peaks} and generated using the state-of-the-art dark-matter-only N-body code \textsc{PkdGrav3}. The cosmological parameters in the used simulations are fixed to the ($\Lambda$CDM,TT,TE,EE+lowE+lensing) results of Planck 2018 \citep{aghanim2020planck}, except for $\Omega_{\mathrm{m}}$ and $\sigma_8$ which are set to the values found in \citet{troxel2018dark}. This setup results in $\Omega_{\mathrm{m}}=0.26$, $\sigma_8=0.84$, $\Omega_{\mathrm{b}}=0.0493$, $n_{\mathrm{s}}=0.9649$, $w=-1$ and $h=0.6736$. We include three massive neutrino species in all simulations. The neutrinos are modelled as a relativistic fluid \citep{tram2019fully} and a degenerate mass hierarchy with a minimal neutrino mass of $m_{\nu}=0.02$ eV per species was chosen. The dark energy density $\Omega_{\Lambda}$ is adapted for each cosmology to achieve a flat geometry. \\
\noindent Each simulation was run using a unit box with a side-length of 900 Mpc/$h$ and $768^3$ simulated particles. In order to achieve a simulation volume large enough to cover the redshift range up to $z=3.0$ the unit box was replicated up to 14 times per dimension depending on the cosmology. While such a replication scheme is known to underpredict the variance of very large, super-box modes \citep{fluri2019cosmological}, it has been demonstrated by \citet{zuercher2021peaks} that the simulations accurately recover the angular power spectra predicted by the theory code \texttt{CLASS} \citep{lesgourgues2011cosmic} for $\ell \in [30, 2048]$. \\

\noindent The particle shells from each \texttt{PKDGRAV3} simulation are combined into tomographic full-sky mass maps using the \texttt{UFalcon} software \citep{sgier2019fast}. The particle shells are weighted according to the tomographic redshift distributions shown in Figure~\ref{fig:redshift_distribution}. The \texttt{UFalcon} software uses the \texttt{HEALPIX}  \citep{gorski2005healpix} pixelization scheme to pixelize the sphere. A resolution of $\texttt{NSIDE} = 1024$ was chosen. \texttt{UFalcon} also makes use of the Born approximation, which is known to deteriorate the accuracy of the produced mass maps. However, \citet{petri2017validity} have demonstrated that the introduced bias is negligible for stage \uppercase\expandafter{\romannumeral3}-like and stage \uppercase\expandafter{\romannumeral4}-like surveys. \\

\noindent The spherical Kaiser-Squires mass mapping technique \citep{kaiser1993mapping, wallis2017mapping} is used to obtain the cosmic shear signal from the simulated mass maps. To forward-model a realistic weak lensing survey a shape noise signal must then be added to the cosmic shear signal and an appropriate survey mask must be applied. The survey masks are chosen such that we obtain eight stage \uppercase\expandafter{\romannumeral3} surveys and two stage \uppercase\expandafter{\romannumeral4} surveys from each full-sky map. \\
\noindent The shape noise signal is obtained in the same way as described in \citet{zurcher2021cosmological}. We randomly sample galaxy positions within the survey region until the target source density is reached. The intrinsic ellipicities of the galaxies are then drawn from a probability distribution that was fit to the observed galaxy ellipticities in \citet{troxel2018dark} (see \citet{zurcher2021cosmological}).
The ellipticity of each individual galaxy is rotated by a random phase.
Using five and twenty shape noise realisations per survey patch, we achieve the desired number of $N=2000$ survey realisations for the stage \uppercase\expandafter{\romannumeral3} and stage \uppercase\expandafter{\romannumeral4} survey setup, respectively. \\

\begin{figure*}
\centering
\includegraphics[width=18cm,height=7cm]{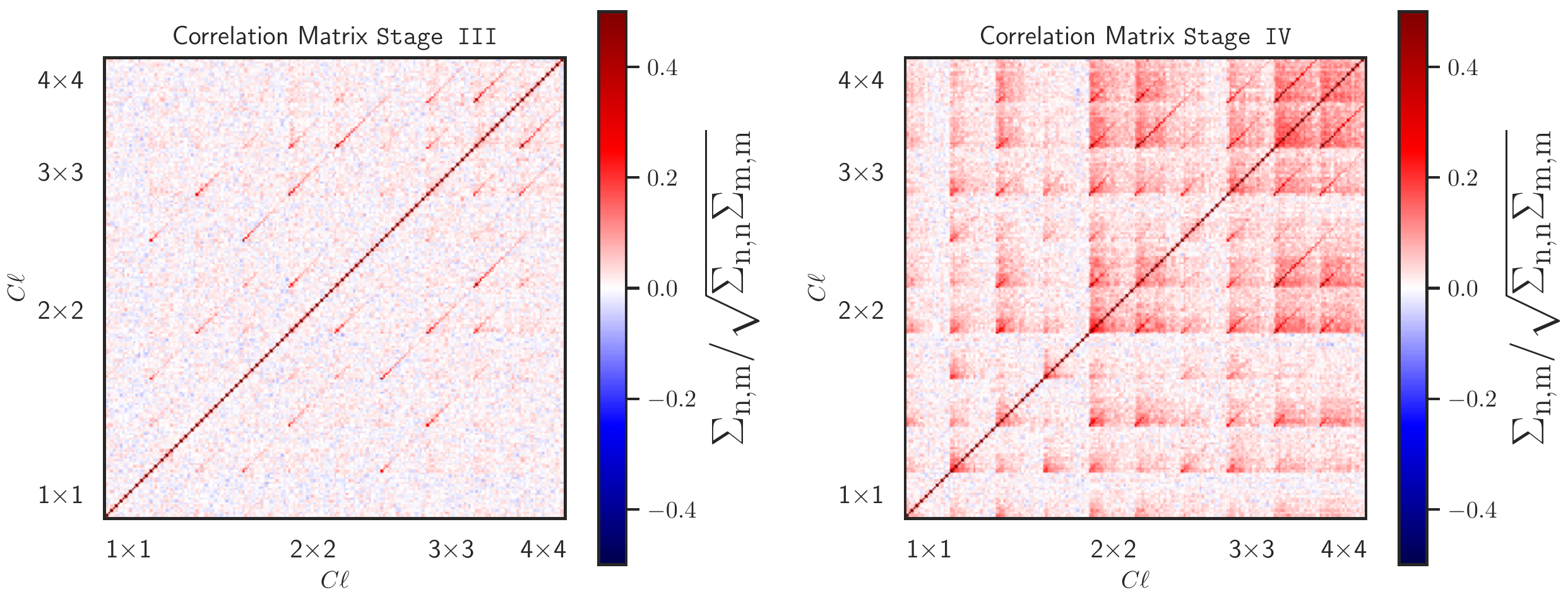}
\caption{Correlation matrices for the stage \uppercase\expandafter{\romannumeral3} survey (left panel) and the stage \uppercase\expandafter{\romannumeral4} survey(right panel). The ordering of the redshift tomographic bin combinations for the angular power spectra
is $1\times1$, $1\times2$, $1\times3$, $1\times4$, $2\times2$, $2\times3$, $2\times4$, $3\times3$, $3\times4$
and $4\times4$, from left to right. For each angular power spectrum, all 20 bins ranging from $\ell$ = 100 to $\ell$ = 1000 are shown.}
\label{fig:Covariance_matrix}
\end{figure*}

\noindent The tomographic angular power spectra realisations $C_{\ell, \mathrm{i}}$ are then measured from the forward-modelled surveys using the \texttt{anafast} routine of the \texttt{healpy} software \citep{zonca2019healpy} using 20 bins from $\ell_{\rm{min}}=100$ to $\ell_{\rm{max}}=1000$, the same as \citet{sgier2019fast}, where the index $i$ runs over the number of survey realisations $N$. The covariance matrix $\Sigma$ is estimated according to  
\begin{equation}
\hat{\Sigma} = \frac{1}{N - 1} \sum_{\mathrm{i}=1}^{N} (C_{\ell, \mathrm{i}} - \bar{C_{\ell}}) (C_{\ell, \mathrm{i}} - \bar{C_{\ell}})^T,
\end{equation}
where $\bar{C_{\ell}}$ indicates the mean of the angular power spectra realisations $C_{\ell, \mathrm{i}}$. The estimated correlation matrices $C_{\mathrm{n,m}} \equiv \Sigma_{\mathrm{n,m}}/\sqrt{\Sigma_{\mathrm{n,n}} \Sigma_{\mathrm{m,m}}}$ are presented in Figure~\ref{fig:Covariance_matrix}.

\subsection{Likelihood Analysis}\label{3.3}
We use a Bayesian likelihood approach to evaluate the cosmological parameter constraints of different predictors. We assume a Gaussian error model and the likelihood is realized by:
\begin{equation}
\begin{aligned}
    &\log\mathcal{L} \\&= -\frac{1}{2}\sum_{ij}(C_{\ell,\rm{truth}}^{i}-C_{\ell,\rm{compare}}^{i})^T\hat{\Sigma}^{-1}(C_{\ell,\rm{truth}}^{j}-C_{\ell,\rm{compare}}^{j})
\end{aligned}
\end{equation}
Here $C_{\ell,\rm{truth}}$ stands for the value of the observable, computed using \texttt{PyCosmo} \citep{refregier2017pycosmo,tarsitano2020predicting, moser2022symbolic} with a chosen predictor and the fiducial cosmological parameters, measured by the Wilkinson Microwave Anisotropy Probe satellite (WMAP) 9 \citep{Hinshaw_2013}, presented in Table \ref{tab:cosmology parameters}. $C_{\ell,\rm{compare}}$ is predicted using another predictor for comparison. The cosmology for the observable is different from what is used for the covariance matrix. However, this effect is neglected assuming the covariance matrix parameter independent \citep{kodwani2018effect}. $\Sigma^{-1}$ is the unbiased estimate of the inverse covariance matrix \citep{cov1,cov2} represented as:
\begin{equation}
    \hat{\Sigma}^{-1} = \frac{N-N^{'}-2}{N-1}\hat{\Sigma}^{-1},
\end{equation}
$N$ is the number of realisations generated from the simulations and $N^{'}$ is the total number of data bins, which is given by
\begin{equation}
    N^{'} = N_{\rm{redshift}}\times N_{\ell}.
\end{equation}
Here we have $N=2000$, $N_{\ell}=20$ and $N_{\rm{redshift}}=10$.
\newline
\begin{table}
\caption{The fiducial values for the cosmological parameters and the flat priors for the cosmological parameters that are varied in the analysis.}
\begin{center}
\begin{tabular}{cccc}
\hline
Parameters\quad&Fiducial \quad&Priors\quad&Priors
\\
&values&(stage-\uppercase\expandafter{\romannumeral3} survey)&(stage-\uppercase\expandafter{\romannumeral4} survey)
\\
\hline
$\Omega_{\rm{m}}$&0.291&$[0,0.6]$&$[0.2,0.4]$\\
\hline
$n_{\rm{s}}$&0.969&$[0.3,2.0]$&$[0.7,1.2]$\\
\hline
$h$&0.69&$[0.1,2.5]$&$[0.4,0.9]$\\
\hline
$\sigma_8$&0.826&$[0.3,1.4]$&$[0.7,0.95]$\\
\hline
$w_0$&-1.0&$[-3.5,0.5]$&$[-2.5,0.5]$\\
\hline
$\Omega_{\rm{b}}$&0.0473&\\
\hline
\end{tabular}
\label{tab:cosmology parameters}
\end{center}
\end{table}
\subsection{Parameter Inference}
The posterior is sampled efficiently using the Markov Chain Monte Carlo (MCMC) ensemble sampler, $\texttt{emcee}$ \citep{Foreman_Mackey_2013}. We vary 4 cosmological parameters $\{\Omega_{\rm{m}},\sigma_8,n_{\rm{s}},h\}$ for the $\Lambda \texttt{CDM}$ cosmological model and an additional parameter $w_0$ for the extended $\texttt{wCDM}$ model, where we fix $w_{\rm{a}}\equiv0$. Table~\ref{tab:cosmology parameters} shows the priors used for these parameters. We run the MCMC chains with $100$ walkers per parameter and cut the burn in phase for each run as one third of the chain length. Each individual chain has more than 100,000 samples. For the visualisation of the marginalised posteriors, we use the public $\texttt{Getdist}$ \citep{lewis2019getdist}.

\section{Results}\label{sec4}
We present the results of our comparison of different predictors in this section, including the analysis of the matter power spectrum, the weak lensing power spectrum, and the cosmological parameter constraints based on the stage \uppercase\expandafter{\romannumeral3} and stage \uppercase\expandafter{\romannumeral4} weak lensing surveys.
\subsection{Power Spectrum}\label{comparison of power spectrum}
\begin{figure*}%[！htpb]/[H]
\centering
\includegraphics[width=18cm,height=9cm]{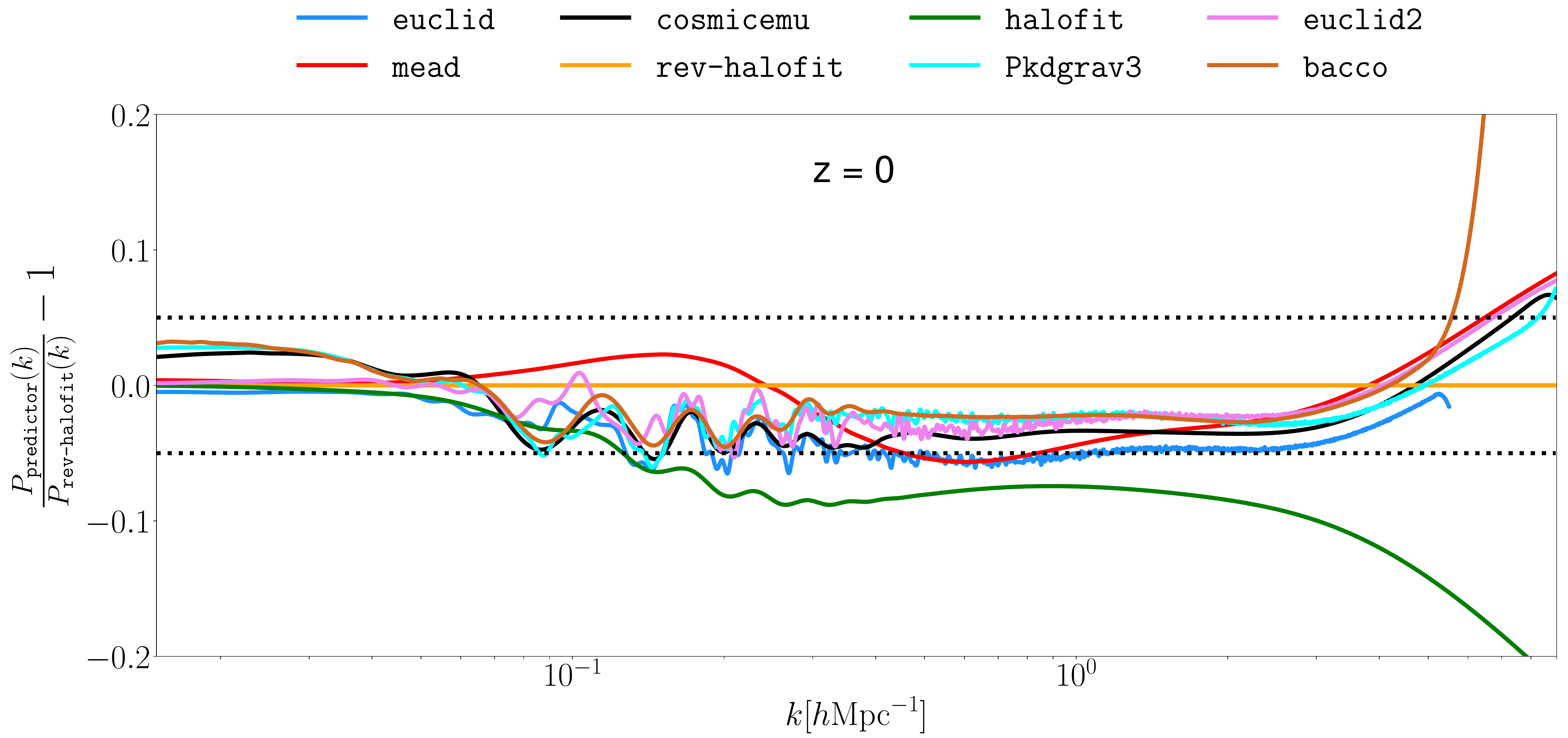}
\caption{Comparison of dark-matter-only, non-linear $P(k)$ predictions for different predictors at redshift $z=0$, subtracted and divided by $\texttt{rev-halofit}$ as reference.}
\label{fig:power spectrum comparison1}
\end{figure*}
We use the linear power spectrum predicted by $\texttt{PyCosmo}$ and generated following \cite{Eisenstein_1999} as the input for all predictors.
Figure~\ref{fig:power spectrum comparison1} shows the comparison of dark-matter-only non-linear $P(k)$ predictions from different predictors at redshift $z=0$, and the comparison for different redshifts ranging from $z=0$ to $z=5$ in Appendix~\ref{sec:Appendix}. The results are shown for $k$ ranging from $k=0.01h\text{Mpc}^{-1}$ to $9h\text{Mpc}^{-1}$ using $10000$ bins. $\texttt{BaccoEmulator}$ and $\texttt{CosmicEmulator}$ are not valid for $z>3$, so we do not present their comparison for the higher redshift at $z=5$. Figure~\ref{fig:power spectrum comparison1} and Figure~\ref{fig:power spectrum comparison2} indicate that:
\begin{itemize}
\item All the predictors except for \texttt{halofit} are within the $5\%$ level of accuracy compared to $\texttt{rev-halofit}$ for $z<2$ and $k<7h\text{Mpc}^{-1}$ (\texttt{BaccoEmulator} is valid for $z<1.5$ and $k<5h\text{Mpc}^{-1}$, see the details in Figure~\ref{fig:power spectrum comparison2}). Note that this is consistent with the comparison of $\texttt{mead}$, $\texttt{rev-halofit}$ and $\texttt{halofit}$ in \citet{10.1093/mnras/stv2036}.
\item $\texttt{halofit}$ shows stronger discrepancies compared with the other predictors at small scales for $k>0.1h\text{Mpc}^{-1}$ and this discrepancy can reach $20\%$ for $k\sim10h\text{Mpc}^{-1}$. 
\item $\texttt{mead}$ and $\texttt{rev-halofit}$ show close agreement with the emulators at the $5\%$ level for $k<9h\text{Mpc}^{-1}$ and $z<0.5$. However, at higher redshifts $1<z<5$, the discrepancies between $\texttt{mead}$ and the emulators can reach $10\%$ for $k>3h\text{Mpc}^{-1}$, while $\texttt{rev-halofit}$ provides a more consistent precision within $5\%$. 
\item All the emulators yield an agreement within the $2-3\%$ level compared with the $\texttt{Pkdgrav3}$ simulation for $k<9h\text{Mpc}^{-1}$ and $z<1.5$. However, this is not valid at higher redshifts. 
\item For large scales with $k<0.5h\text{Mpc}^{-1}$, the different predictors show a better agreement at higher redshifts.
\end{itemize}
\subsection{Weak Lensing Power Spectrum}\label{sec:4.2}
\begin{figure*}%[！htpb]/[H]
\centering
\includegraphics[width=18cm,height=12cm]{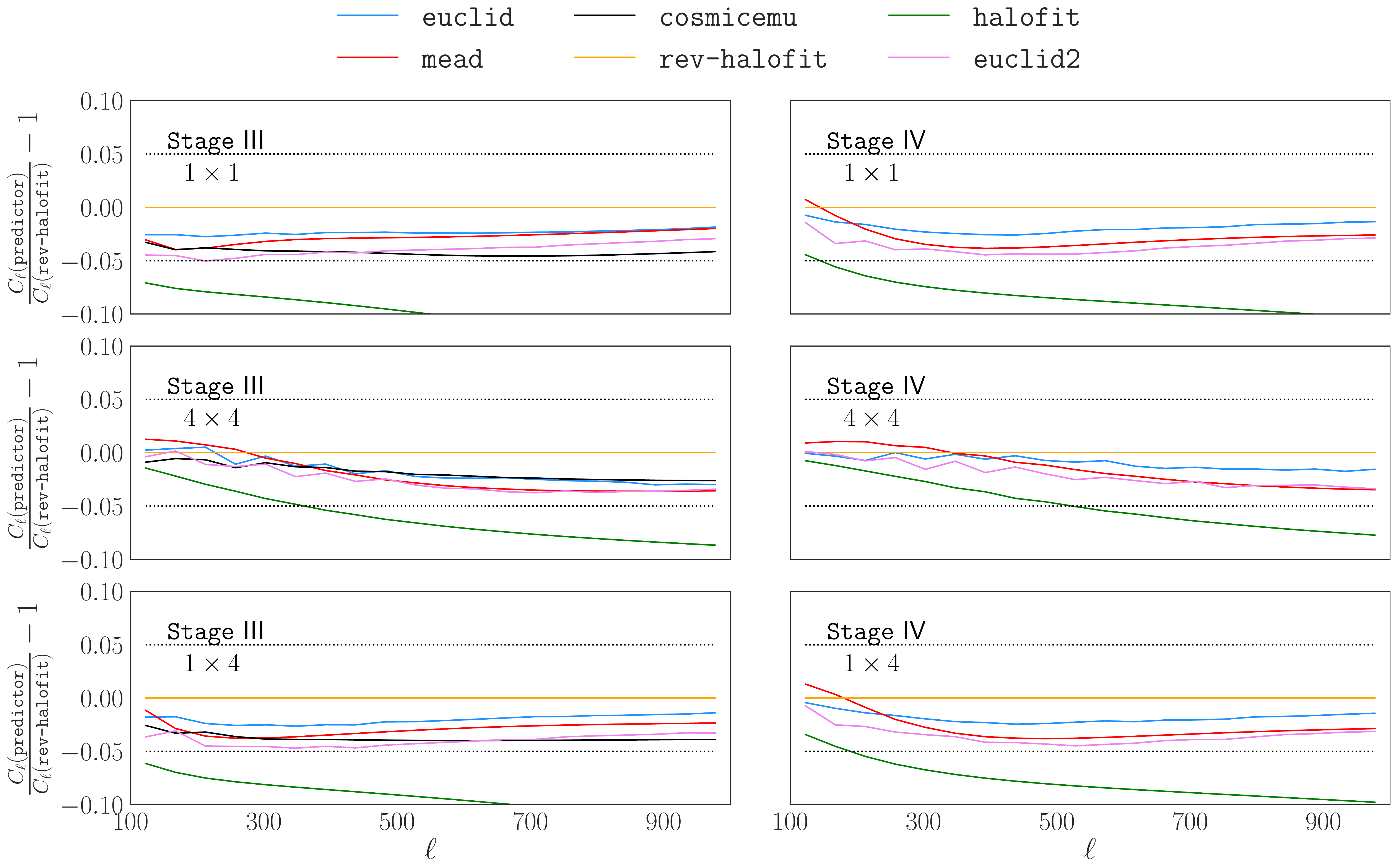}
\caption{The comparison of weak lensing shear $C_\ell$s for different predictors. Each $C_\ell$ is multiplied by $\ell(\ell+1)/2\pi$. The upper $2$ panels in each column show the auto-correlated $C_\ell$s for the first, and the fourth redshift bin and the bottom ones shows the cross correlated $C_\ell$s between these two bins. The left-hand panels show the plots for the Stage \uppercase\expandafter{\romannumeral3} survey and the right-hand side shows the Stage \uppercase\expandafter{\romannumeral4} survey results.}
\label{fig: Cl comparison}
\end{figure*}
We compute the weak lensing shear power spectrum $C_\ell$ for the Stage \uppercase\expandafter{\romannumeral3} survey and the Stage \uppercase\expandafter{\romannumeral4} survey with different predictors. Limited by the range of $k_{\rm{max}}$ of the emulators, the $C_\ell$s are computed using $20$ $\ell$-bins spaced linearly between $\ell_{\rm{min}}=100$ and $\ell_{\rm{max}}=1000$. The integrated redshift range is $[0.08,2.0]$ for the stage \uppercase\expandafter{\romannumeral3} survey and $[0.08,3.0]$ for the stage \uppercase\expandafter{\romannumeral4} survey. This setting was chosen in order to avoid the instability of emulators for low redshifts, where we found that $\texttt{EuclidEmulator}$ and $\texttt{EuclidEmulator2}$ predict the $C_\ell$s with a discrepency larger than $10\%$ at $z<0.08$. This choice differs from the setting used for the generation of the covariance matrix. However, we find that this only changes the discrepencies between different predictors for $C_\ell$s by $0.1\%$, since only $1\%$ of the low-redshift galaxies are missed for the stage \uppercase\expandafter{\romannumeral3} survey and $0.1\%$ of the galaxies for the stage \uppercase\expandafter{\romannumeral4} survey. Using this redshift range, we have to exclude $\texttt{CosmicEmulator}$ from the comparison for the stage \uppercase\expandafter{\romannumeral4} survey as it allows only up to $z=2.0$. The comparison is shown in Figure~\ref{fig: Cl comparison}, with the left-hand panels showing the results for the Stage \uppercase\expandafter{\romannumeral3} survey and the right-hand side showing the Stage \uppercase\expandafter{\romannumeral4} survey results. In the individual panels, we present $C_{\ell}\, \ell(\ell+1)/2\pi$ for each predictor and illustrate the comparison by subtracting and dividing $\texttt{rev-halofit}$ as the reference. In Figure~\ref{fig: Cl comparison} the first row shows the comparison of the auto-correlated $C_\ell$s for the redshift bins $1\times1$, the second row for $4\times4$, and the bottom row shows the cross correlated $C_\ell$s for $1\times4$. From Figure~\ref{fig: Cl comparison}, one can infer that: \begin{itemize}
\item All the predictors, except for $\texttt{halofit}$, yield an agreement at the $5\%$ level, both for the auto and cross $C_\ell$. This is consistent with our results for $P(k)$. 
\item $\texttt{mead}$ shows a good agreement with $\texttt{CosmicEmulator}$, $\texttt{EuclidEmulator2}$ and $\texttt{EuclidEmulator}$, while $\texttt{rev-halofit}$ exhibits a larger discrepancy. 
\item The comparison of $C_\ell$ for different predictors does not show a significant difference between the stage \uppercase\expandafter{\romannumeral3} survey and the stage \uppercase\expandafter{\romannumeral4} survey.
\end{itemize}
\subsection{Cosmological Parameters Constraints}
The comparison of the weak lensing cosmological parameter constraints for different predictors is present in this section. As indicated in Section~\ref{sec:3}, we consider a stage \uppercase\expandafter{\romannumeral3} survey and a stage \uppercase\expandafter{\romannumeral4} survey. For each survey, we perform a comparison using the standard $\Lambda \text{CDM}$ cosmological model and the extended $w\text{CDM}$ model. A summary of the constraints on $\{S_8,\Omega_{\rm{m}},w_0\}$ is presented in Table~\ref{tab:S8 deviation all}, and the constraints on $\{S_8,\Omega_{\rm{m}},n_{\rm{s}},h,w_0\}$ in Table~\ref{tab:S8 deviation all2}.
\subsubsection{$\Lambda \text{CDM}$ cosmology constraints}

\begin{figure*}%[！htpb]/[H]
\centering
\includegraphics[width=18cm,height=18cm]{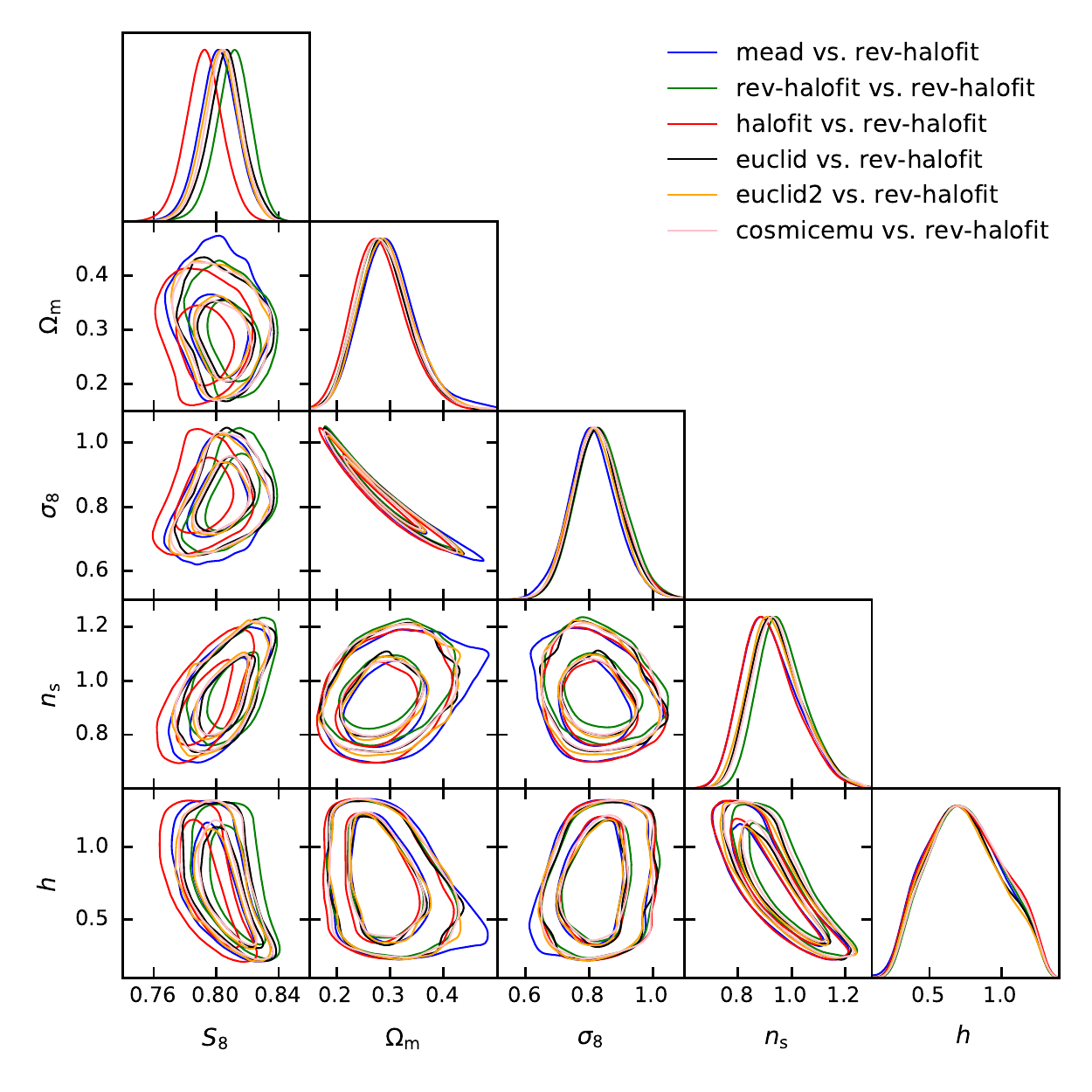}
\caption{Cosmological parameter constraints for the stage \uppercase\expandafter{\romannumeral3} survey in the $\Lambda \texttt{CDM}$ model. For each constraint, $C_{\ell,\rm{truth}}$ is predicted using the first predictor shown in the legend, and $C_{\ell,\rm{compare}}$ computed using the second predictor, as indicated in Section~\ref{3.3}. For the stage \uppercase\expandafter{\romannumeral3} survey, we set $C_{\ell,\rm{truth}}$ with the halo-model based fitting functions ($\texttt{rev-halofit}$, $\texttt{mead}$ and $\texttt{halofit}$) and $3$ emulators ($\texttt{EuclidEmulator}$, $\texttt{EuclidEmulator2}$ and $\texttt{CosmicEmulator}$), and compare with predictions from only the fitting functions (in this figure only $\texttt{rev-halofit}$).}
\label{fig:5params stage3}
\end{figure*}

\begin{figure*}%[！htpb]/[H]
\centering
\includegraphics[width=18cm,height=18cm]{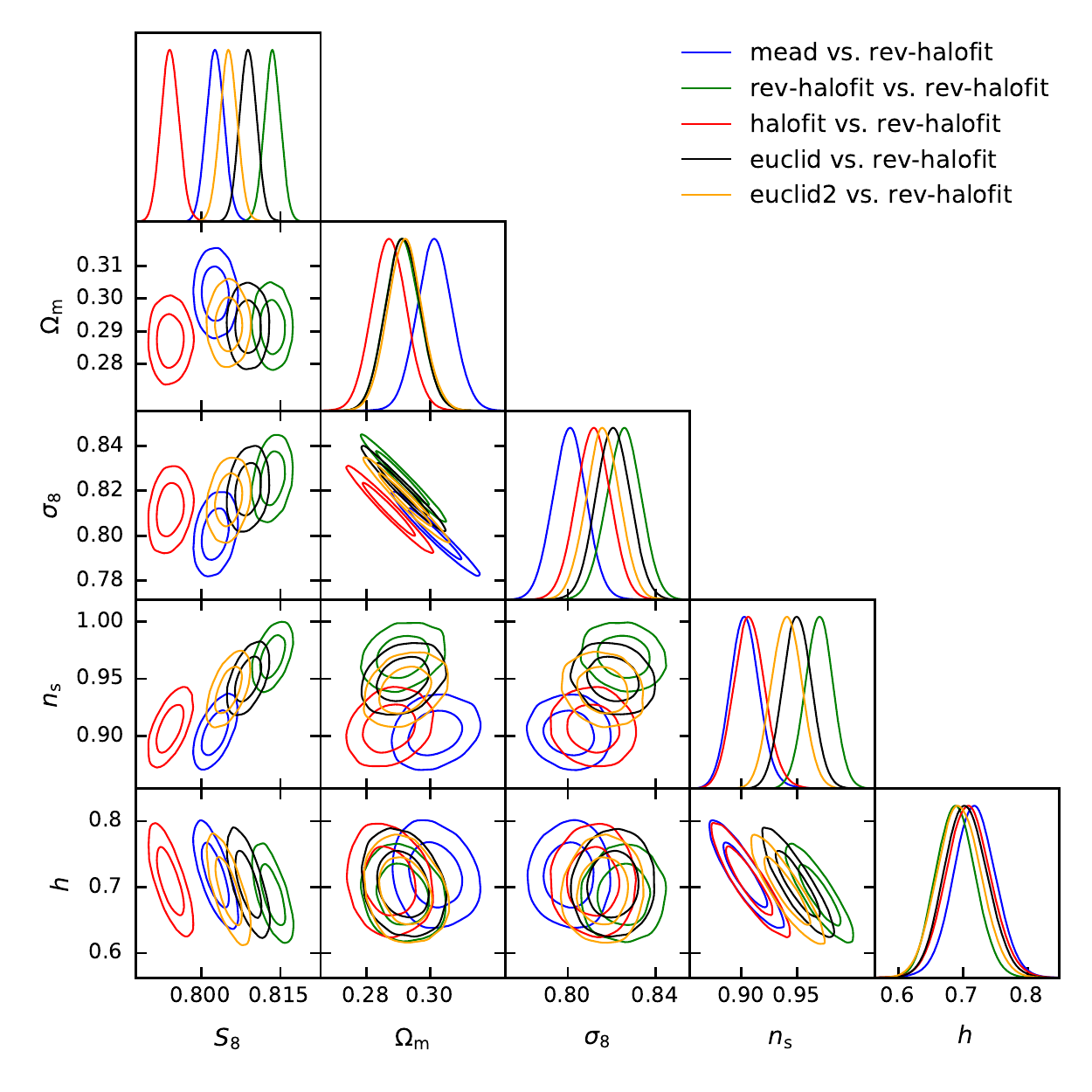}
\caption{Cosmological parameter constraints of the stage \uppercase\expandafter{\romannumeral4} survey in the $\Lambda \texttt{CDM}$ model. Only $2$ emulators, i.e. $\texttt{EuclidEmulator}$ and $\texttt{EuclidEmulator2}$, are chosen for $C_{\ell,\rm{truth}}$, as $\texttt{CosmicEmulator}$ does not provide a sufficient redshift range for the stage \uppercase\expandafter{\romannumeral4} survey.}
\label{fig:5params stage4}
\end{figure*}

\begin{figure*}%[！htpb]/[H]
\centering
\includegraphics[width=10cm,height=22cm]{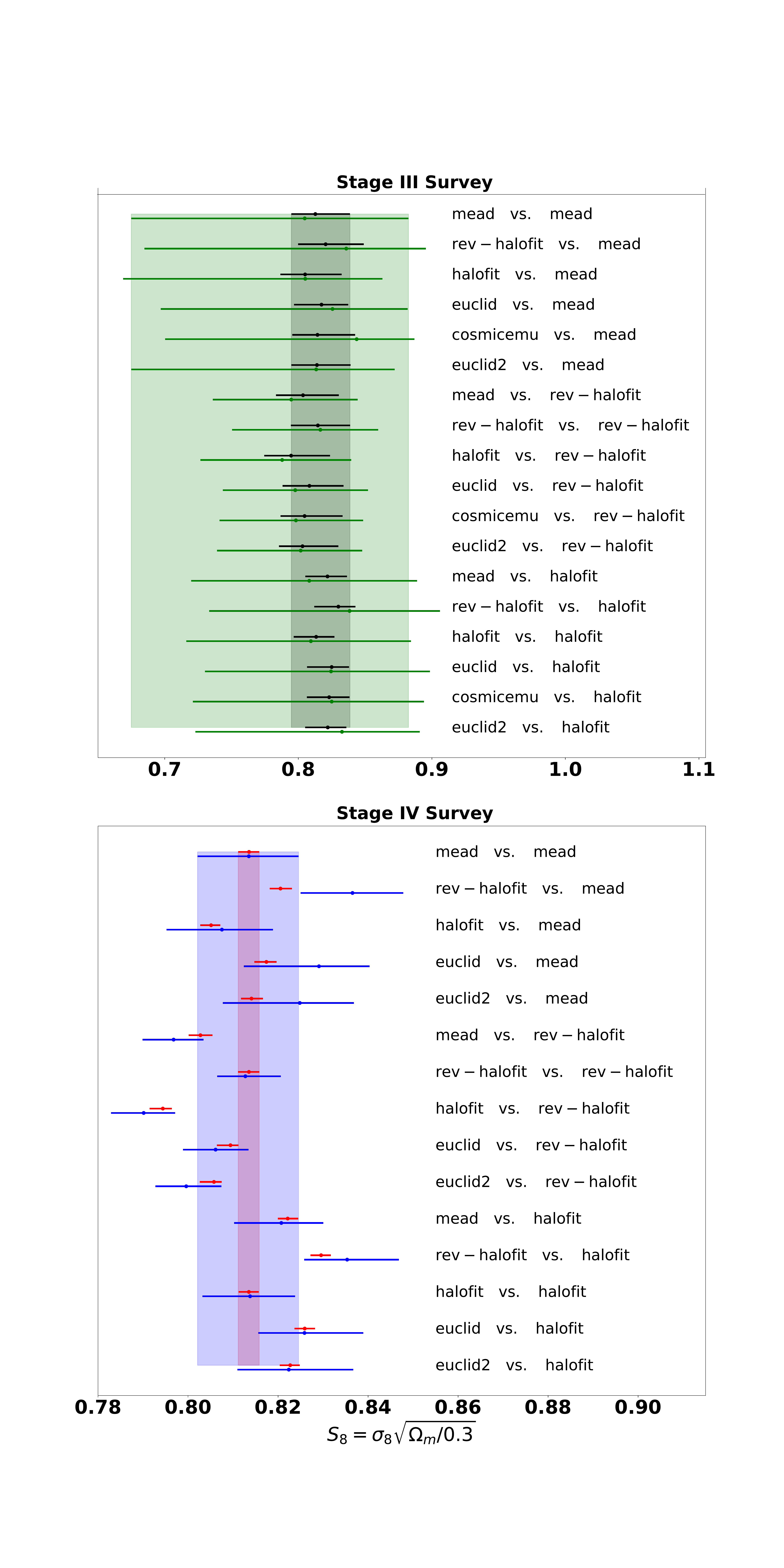}
\caption{Deviations of the parameter constraints on $S_8$. The upper plot shows the result for the stage \uppercase\expandafter{\romannumeral3} survey, for the $\Lambda \texttt{CDM}$ model (black) and the $w \texttt{CDM}$ model (green), respectively. The lower plot shows the stage \uppercase\expandafter{\romannumeral4} survey, for the $\Lambda \texttt{CDM}$ (red) and $w \texttt{CDM}$ (blue), respectively.}
\label{fig:S8 deviation stage3}
\end{figure*}

We present the two-dimensional 68\% and 95\% confidence level contours of the posterior distributions for the $\Lambda \text{CDM}$ model in Figure~\ref{fig:5params stage3} and Figure~\ref{fig:5params stage4} for the Stage 3 and Stage 4 survey setup, respectively. The parameters $\{\Omega_{\rm{m}},\sigma_8,n_{\rm{s}},h\}$ are varied in the MCMC analysis. We additionally compute the constraints on $S_8$, and summarise the shifts in $S_8$ in Figure~\ref{fig:S8 deviation stage3}, presenting the median values of the posteriors and the error bars indicating the 68\% confidence limits of the constraints. One can infer from the posterior distributions in Figure~\ref{fig:S8 deviation stage3} and Table~\ref{tab:S8 deviation all2} that the agreement on $S_8$ between different predictors is less than $0.6\sigma$ for the stage \uppercase\expandafter{\romannumeral3} survey ($0.2-0.3\sigma$ if $\texttt{halofit}$ excluded), while being much larger for the stage \uppercase\expandafter{\romannumeral4} survey. This is caused by the higher constraining power of the \uppercase\expandafter{\romannumeral4} survey. More specifically, the agreements are generally on the $1.4-6.1\sigma$ level ($1.4-3.0\sigma$ if $\texttt{halofit}$ excluded). $\texttt{mead}$ shows good agreement with $\texttt{CosmicEmulator}$, $\texttt{EuclidEmulator}$ and $\texttt{EuclidEmulator2}$ for the stage \uppercase\expandafter{\romannumeral3} survey while it only agrees well with $\texttt{EuclidEmulator2}$ for the stage \uppercase\expandafter{\romannumeral4} survey. The constraints on $h$ do not show significant discrepencies for both surveys, while $n_{\rm{s}}$ reveals discrepencies of several $\sigma$s for different predictors for the stage \uppercase\expandafter{\romannumeral4} survey.
\\
\\
\subsubsection{$\text{wCDM}$ cosmology constraints}
\begin{figure*}%[！htpb]/[H]
\centering
\includegraphics[width=18cm,height=18cm]{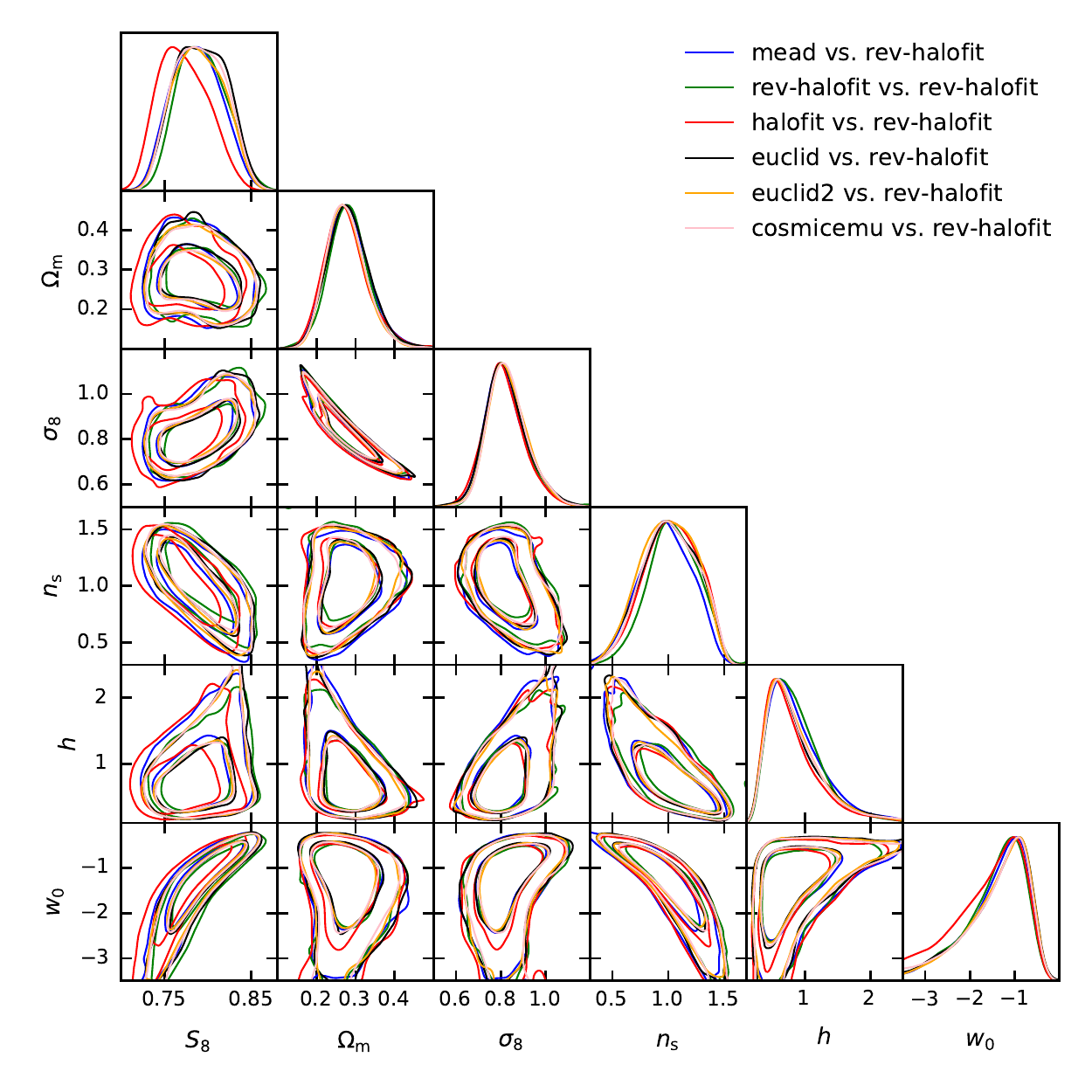}
\caption{Cosmological parameter constraints of the stage \uppercase\expandafter{\romannumeral3} survey in the $\texttt{wCDM}$ cosmological model. Including $w_0$ reduces significantly the constraining power, yielding much broader contours than the $\Lambda \texttt{CDM}$ model.}
\label{fig:6params stage3}
\end{figure*}

\begin{figure*}%[！htpb]/[H]
\centering
\includegraphics[width=18cm,height=18cm]{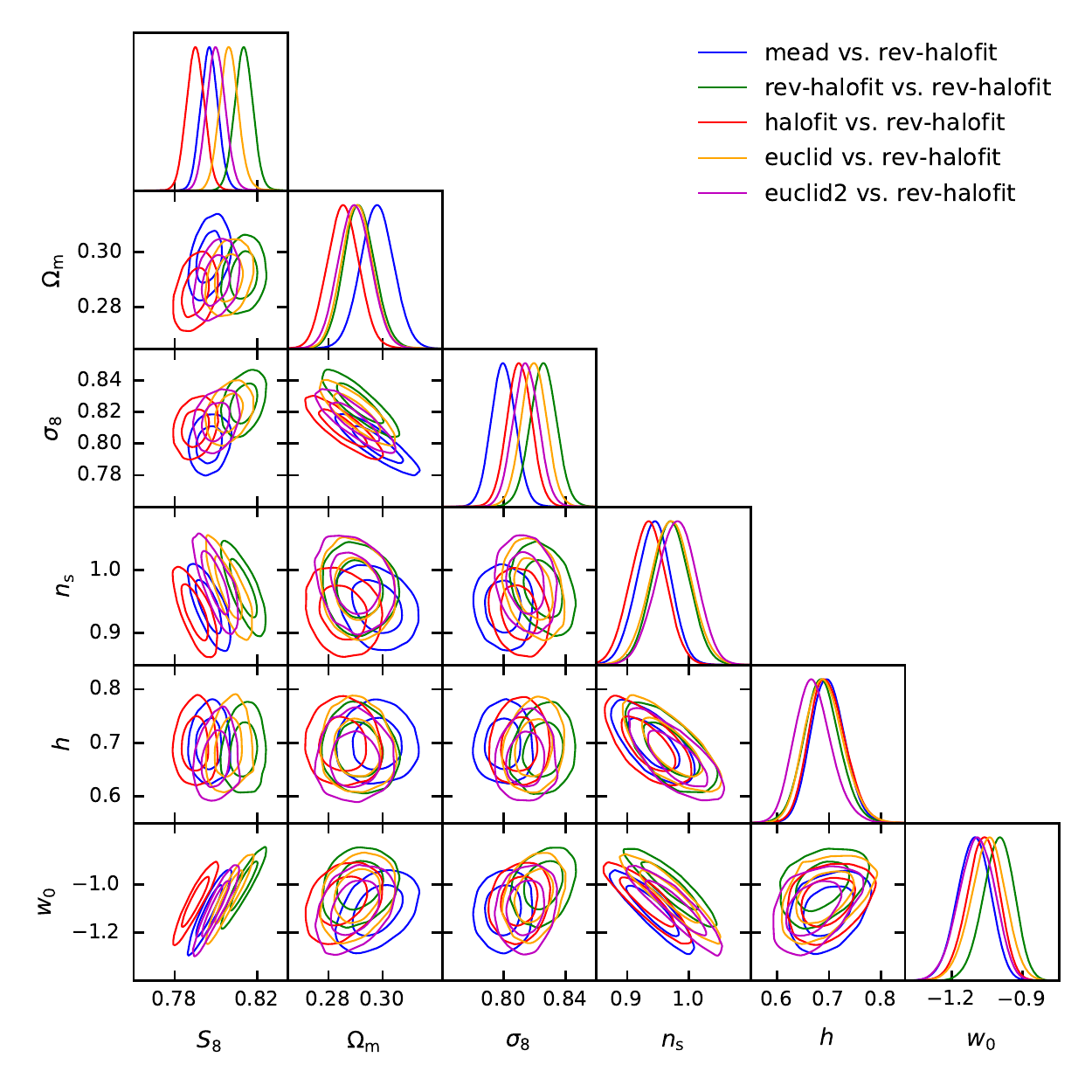}
\caption{Cosmological parameter constraints of the stage \uppercase\expandafter{\romannumeral4} survey in the $w\texttt{CDM}$ cosmological model. The discrepancies between the predictors are alleviated, taking into account a simple $w\texttt{CDM}$ cosmological model with a varying $w_0$.}
\label{fig:6params stage4}
\end{figure*}

We consider the constraining power of weak lensing surveys on dark energy parameters by adopting a time dependent dynamical dark energy equation of state, the CPT-parameterisation \citep{chevallier2001accelerating,PhysRevLett.90.091301}, as an extension to the $\Lambda \text{CDM}$ model. The equation of state parameter is given by
\begin{equation}
    w(a) = w_0 + w_a(1 - a).
\end{equation}
where we use a fixed $w_{\rm{a}}=0$ and a free $w_0$. We present the two-dimensional marginal posterior distributions for the $w\texttt{CDM}$ cosmology parameters in Figure~\ref{fig:6params stage3} and Figure~\ref{fig:6params stage4}, for the stage \uppercase\expandafter{\romannumeral3} survey and the stage \uppercase\expandafter{\romannumeral4} survey, respectively. Taking into account the dark energy model changes the shape and the contour size of the posterior distributions, decreasing the constraining power on the cosmological parameters. The discrepancies in $S_8$ between predictors are generally smaller compared with the $\Lambda\texttt{CDM}$ model due to the decrease in constraining power: $0.18-0.34\sigma$ for the stage \uppercase\expandafter{\romannumeral3} survey and $0.7-2.4\sigma$ for the stage \uppercase\expandafter{\romannumeral3} survey ($0.18-0.26\sigma$ and $0.7-1.7\sigma$ if $\texttt{halofit}$ is excluded, respectively). $\texttt{mead}$ still shows good agreement with  $\texttt{EuclidEmulator}$ and $\texttt{EuclidEmulator2}$ for both the stage \uppercase\expandafter{\romannumeral3} survey and the stage \uppercase\expandafter{\romannumeral4} survey. $\texttt{rev-halofit}$ agrees with all the predictors within $0.3\sigma$ for the stage \uppercase\expandafter{\romannumeral3} survey, and shows discrepancies at the $0.7-2.4\sigma$ level for the stage \uppercase\expandafter{\romannumeral4} survey. 


\begin{table*}%llllllllllllll
\begin{center}
\begin{tabular}{cccccccc}
\hline
Survey & Predictor& $S_8$ &\scriptsize($\sigma$)& $\Omega_{\rm{m}}$&\scriptsize($\sigma$) & $w_0$&\scriptsize($\sigma$) \\
Cosmology&  ref: \texttt{rev-halofit} \\ 
\hline
\multirow{6}{*}{stage-\uppercase\expandafter{\romannumeral3}}& $\texttt{rev-halofit} $&  $0.8147^{+0.0241}_{-0.0203}$&&  $0.288^{+0.0817}_{-0.0662}$&&&\\
& $\texttt{mead} $&  $0.8035^{+0.0269}_{-0.0202}$&  $0.33$&  $0.2996^{+0.0848}_{-0.0698}$&  $0.11$&&\\
\multirow{4}{*}{$\Lambda \texttt{CDM}$}& $\texttt{halofit} $&  $0.7946^{+0.0292}_{-0.0201}$&  $0.57$&  $0.2884^{+0.0783}_{-0.074}$&  $0.0$&&\\
& $\texttt{euclid} $&  $0.8083^{+0.0256}_{-0.0201}$&  $0.2$&  $0.2987^{+0.0831}_{-0.0709}$&  $0.1$&&\\
& $\texttt{cosmicemu} $&  $0.8047^{+0.0285}_{-0.018}$&  $0.29$&  $0.2916^{+0.0789}_{-0.0741}$&  $0.03$&&\\
& $\texttt{euclid2} $&  $0.8031^{+0.0269}_{-0.0177}$&  $0.34$&  $0.2887^{+0.0835}_{-0.0679}$&  $0.01$&&\\
\hline
\multirow{6}{*}{stage-\uppercase\expandafter{\romannumeral3}}& $\texttt{rev-halofit} $&  $0.8165^{+0.0433}_{-0.0661}$&&  $0.2846^{+0.092}_{-0.09}$&&  $-0.9242^{+0.4704}_{-2.294}$&\\
& $\texttt{mead} $&  $0.7947^{+0.0497}_{-0.0588}$&  $0.26$&  $0.31^{+0.0824}_{-0.1022}$&  $0.18$&  $-1.139^{+0.647}_{-2.2626}$&  $0.09$\\
\multirow{4}{*}{$w \texttt{CDM}$}& $\texttt{halofit} $&  $0.7879^{+0.0517}_{-0.0612}$&  $0.34$&  $0.2968^{+0.0787}_{-0.1011}$&  $0.09$&  $-1.1333^{+0.6581}_{-2.3122}$&  $0.09$\\
& $\texttt{euclid} $&  $0.7977^{+0.0545}_{-0.0542}$&  $0.22$&  $0.3049^{+0.085}_{-0.1017}$&  $0.15$&  $-1.1886^{+0.7187}_{-2.1508}$&  $0.11$\\
& $\texttt{cosmicemu} $&  $0.7982^{+0.0504}_{-0.0572}$&  $0.22$&  $0.2931^{+0.0921}_{-0.0969}$&  $0.06$&  $-1.1408^{+0.6926}_{-2.3046}$&  $0.09$\\
& $\texttt{euclid2} $&  $0.8018^{+0.0461}_{-0.0627}$&  $0.18$&  $0.2896^{+0.0928}_{-0.0877}$&  $0.04$&  $-1.0254^{+0.5498}_{-2.2745}$&  $0.04$\\
\hline
\multirow{6}{*}{stage-\uppercase\expandafter{\romannumeral4}}& $\texttt{rev-halofit} $&  $0.8135^{+0.0023}_{-0.0024}$&&  $0.2915^{+0.0077}_{-0.0084}$&&&\\
& $\texttt{mead} $&  $0.8028^{+0.0027}_{-0.0026}$&  $2.96$&  $0.3008^{+0.0094}_{-0.0074}$&  $0.87$&&\\
\multirow{4}{*}{$\Lambda \texttt{CDM}$}& $\texttt{halofit} $&  $0.7944^{+0.002}_{-0.0029}$&  $6.11$&  $0.2856^{+0.0097}_{-0.0064}$&  $0.46$&&\\
& $\texttt{euclid} $&  $0.8094^{+0.0018}_{-0.003}$&  $1.37$&  $0.2917^{+0.0079}_{-0.0084}$&  $0.02$&&\\
& $\texttt{euclid2} $&  $0.8058^{+0.0017}_{-0.0032}$&  $2.62$&  $0.2926^{+0.0079}_{-0.0084}$&  $0.1$&&\\
\hline
\multirow{6}{*}{stage-\uppercase\expandafter{\romannumeral4}}& $\texttt{rev-halofit} $&  $0.8127^{+0.0079}_{-0.0063}$&&  $0.2909^{+0.0095}_{-0.0086}$&&  $-1.0127^{+0.1171}_{-0.1046}$&\\
& $\texttt{mead} $&  $0.7968^{+0.0067}_{-0.0069}$&  $1.73$&  $0.2979^{+0.0106}_{-0.0092}$&  $0.53$&  $-1.106^{+0.1107}_{-0.1163}$&  $0.61$\\
\multirow{4}{*}{$w \texttt{CDM}$}& $\texttt{halofit} $&  $0.7902^{+0.007}_{-0.0073}$&  $2.39$&  $0.2856^{+0.0093}_{-0.0096}$&  $0.42$&  $-1.0646^{+0.1069}_{-0.1197}$&  $0.35$\\
& $\texttt{euclid} $&  $0.8061^{+0.0073}_{-0.0072}$&  $0.68$&  $0.2908^{+0.0088}_{-0.0094}$&  $0.01$&  $-1.046^{+0.1142}_{-0.1288}$&  $0.22$\\
& $\texttt{euclid2} $&  $0.7996^{+0.0078}_{-0.0069}$&  $1.31$&  $0.2901^{+0.0099}_{-0.0094}$&  $0.06$&  $-1.0965^{+0.1288}_{-0.1255}$&  $0.51$\\
\hline

\end{tabular}
\caption{Numerical constraints on the cosmological parameters corresponding to the contours in Figure~\ref{fig:5params stage3}, \ref{fig:5params stage4}, \ref{fig:6params stage3}, and \ref{fig:6params stage4}. For each predictor, the $\sigma$s show the theoretical discrepancies for each parameter, compared to the reference one.}
\label{tab:S8 deviation all}
\end{center}
\end{table*}

\section{Conclusions}\label{sec5}
The different halo-model based fitting functions and emulators have been widely used for the prediction of non-linear power spectrum to study the large scale structure of the Universe. It is essential to understand their advantages, limitations, and theoretical uncertainties for different surveys and cosmologies. From our results, we conclude that:
\begin{itemize}
\item Compared with \texttt{Pkdgrav3} simulations, the halo-model based fitting functions, except $\texttt{halofit}$, yield a $5-10\%$ level accuracy for the matter power spectrum $P(k)$ for $k<9h\text{Mpc}^{-1}$ and $z<2$, while emulators show better precision at the $2\%$ level. For the weak lensing shear power spectrum $C_\ell$, all the predictors, except for $\texttt{halofit}$, show a $5\%$ level mutual agreement.
\item For the stage \uppercase\expandafter{\romannumeral3} survey with a $\Lambda \texttt{CDM}$ cosmology, the agreement on $S_8$ between different predictors are within $0.6\sigma$, and within $0.2\sigma$ for other cosmological parameters ($0.3\sigma$ and $0.2\sigma$ if we exclude $\texttt{halofit}$, respectively). This indicates the applicability of the studied predictors for the stage \uppercase\expandafter{\romannumeral3} surveys.
\item For the stage \uppercase\expandafter{\romannumeral4} survey using a $\Lambda \texttt{CDM}$ cosmology, the disagreements on $S_8$ are increased to several $\sigma$s,  with the largest discrepancy of $6.1\sigma$ between \texttt{rev-halofit} and $\texttt{halofit}$, and the best agreement between \texttt{mead} and \texttt{EuclidEmulator2}.
\item If $w_0$ is taken into account for the $w\texttt{CDM}$ cosmology, we get weaker constraints on $S_8$, and the discrepancies between different predictors are reduced to $0.2-0.3\sigma$ and $0.7-2.4\sigma$ for the stage \uppercase\expandafter{\romannumeral3} survey and the stage \uppercase\expandafter{\romannumeral4} survey respectively ($0.18-0.26\sigma$ and $0.7-1.7\sigma$ if we exclude $\texttt{halofit}$, respectively).
\item The accuracy of the current fitting function models and emulators therefore appear sufficient for stage \uppercase\expandafter{\romannumeral3} surveys. However, for the future \uppercase\expandafter{\romannumeral4} surveys, our results suggest that the fitting function models are currently not sufficiently accurate, and would need further improvements in the future. For emulators, it is required to explore wider ranges of cosmological parameters, $k$-modes, and redshifts, while pursuing consistent precision with reliable hydrodynamic N-body simulations. 
\end{itemize} 

\noindent Note that, in this study, we include dark-matter-only predictions, without any consideration of baryonic effects, which can have a strong impact on small scales \citep{jing2006influence,rudd2008effects}. Current studies of halo-model based fitting functions already include other systematics, i.e. massive neutrino and baryonic effects like AGN feedback and gas cooling. The inclusion of these systematics will significantly reduce the constraining power, and might alleviate the discrepancies between the predictors. There are also other sources of uncertainties in weak lensing experiments that we did not include in this work and that could affect the our results, e.g. photometric redshift uncertainty \citep{hildebrandt2020kids+,huterer2006systematic,choi2016cfhtlens}, shear bias \citep{bernstein2002shapes,hirata2004galaxy,bernstein2010shape,refregier2012noise,melchior2012means} and galaxy intrinsic alignment \citep{heavens2000intrinsic,fluri2019cosmological,hirata2004intrinsic,bridle2007dark,joachimi2011constraints}.

\section*{Acknowledgments}
%\addcontentsline{toc}{section}{Acknowledgement}
This work was supported in part by grant 200021\_192243 from the Swiss National Science Foundation.
\newline
We thank Mischa Knabenhans from University of Z\"urich for the distribution of $\texttt{Pkdgrav-3}$. We further thank Aurel Schneider from the University of Z\"urich for the useful discussions regarding this project and the covariance matrix for a stage \uppercase\expandafter{\romannumeral4} survey.
We would also like to thank Uwe Schmitt from ETH Z\"urich for his support with the GitLab server and development of $\texttt{PyCosmo}$.
\newline
The Collaborating Institutions are the Eidgen\"ossische Technische
Hochschule (ETH) Z\"urich, Ecole Polytechnique, the Laboratoire de Physique Nucl\'eaire et des Hautes Energies of Sorbonne University. 
\newline
\section*{Data Availability}
Most of the analysis in this work is down on the Euler cluster\footnote{\url{https://scicomp.ethz.ch/wiki/Euler}} operated by ETH Zurich. Here follows the computational codes used in this study: $\texttt{PyCosmo}$ \citep{refregier2017pycosmo,tarsitano2020predicting,moser2022symbolic} is used as the main tool where all the non linear codes are implemented for the computation of auto (cross) power spectra, galaxy reshift distribution counts, and observable of cosmic shear. It is also extended to include interfaces with the emulators. $\texttt{Anafast}$ is used for computation of power spectra from simulations, and all the the maps (masks, weight, shear, mass) in pipeline are in $\texttt{HealPix}$ format. We use $\texttt{Emcee-3.0.2}$ \citep{Foreman_Mackey_2013} for the sampling of parameter space and $\texttt{Getdist}$ \citep{lewis2019getdist} for the plotting of likelihood contours and $\texttt{Uhammer}$ for the simplification of $\texttt{Emcee}$ running.
Some of the results in this paper have been derived using the $\texttt{healpy}$ and $\texttt{HEALPix packages}$
\citep{gorski1999healpix}.
In this study, we made use of the functionalities provided by $\texttt{numpy}$ \citep{numpy}, $\texttt{scipy}$ \citep{Virtanen2020Mar} and
$\texttt{matplotlib}$ \citep{matplotlib}.
\newpage

%%%%%%%%%%%%%%%%%%%% REFERENCES %%%%%%%%%%%%%%%%%%

\bibliographystyle{mnras}
\bibliography{refrence} 

\begin{thebibliography}{}
\makeatletter
\relax
\def\mn@urlcharsother{\let\do\@makeother \do\$\do\&\do\#\do\^\do\_\do\%\do\~}
\def\mn@doi{\begingroup\mn@urlcharsother \@ifnextchar [ {\mn@doi@}
  {\mn@doi@[]}}
\def\mn@doi@[#1]#2{\def\@tempa{#1}\ifx\@tempa\@empty \href
  {http://dx.doi.org/#2} {doi:#2}\else \href {http://dx.doi.org/#2} {#1}\fi
  \endgroup}
\def\mn@eprint#1#2{\mn@eprint@#1:#2::\@nil}
\def\mn@eprint@arXiv#1{\href {http://arxiv.org/abs/#1} {{\tt arXiv:#1}}}
\def\mn@eprint@dblp#1{\href {http://dblp.uni-trier.de/rec/bibtex/#1.xml}
  {dblp:#1}}
\def\mn@eprint@#1:#2:#3:#4\@nil{\def\@tempa {#1}\def\@tempb {#2}\def\@tempc
  {#3}\ifx \@tempc \@empty \let \@tempc \@tempb \let \@tempb \@tempa \fi \ifx
  \@tempb \@empty \def\@tempb {arXiv}\fi \@ifundefined
  {mn@eprint@\@tempb}{\@tempb:\@tempc}{\expandafter \expandafter \csname
  mn@eprint@\@tempb\endcsname \expandafter{\@tempc}}}

\bibitem[\protect\citeauthoryear{Aghanim et~al.,}{Aghanim
  et~al.}{2020}]{aghanim2020planck}
Aghanim N.,  et~al., 2020, Astronomy \& Astrophysics, 641, A6

\bibitem[\protect\citeauthoryear{Angulo, Springel, White, Jenkins, Baugh  \&
  Frenk}{Angulo et~al.}{2012}]{LGadget3}
Angulo R.~E.,  Springel V.,  White S. D.~M.,  Jenkins A.,  Baugh C.~M.,   Frenk
  C.~S.,  2012, \mn@doi [Monthly Notices of the Royal Astronomical Society]
  {10.1111/j.1365-2966.2012.21830.x}, 426, 2046

\bibitem[\protect\citeauthoryear{Angulo, Zennaro, Contreras, Aricò,
  Pellejero-Ibañez  \& Stücker}{Angulo et~al.}{2020}]{angulo2020bacco}
Angulo R.~E.,  Zennaro M.,  Contreras S.,  Aricò G.,  Pellejero-Ibañez M.,
  Stücker J.,  2020, The BACCO Simulation Project: Exploiting the full power
  of large-scale structure for cosmology (\mn@eprint {arXiv} {2004.06245})

\bibitem[\protect\citeauthoryear{Aric{\`o}, Angulo, Contreras, Ondaro-Mallea,
  Pellejero-Iba{\~n}ez  \& Zennaro}{Aric{\`o} et~al.}{2021}]{arico2021bacco}
Aric{\`o} G.,  Angulo R.~E.,  Contreras S.,  Ondaro-Mallea L.,
  Pellejero-Iba{\~n}ez M.,   Zennaro M.,  2021, Monthly Notices of the Royal
  Astronomical Society, 506, 4070

\bibitem[\protect\citeauthoryear{Bartelmann \& Maturi}{Bartelmann \&
  Maturi}{2016}]{bartelmann2016weak}
Bartelmann M.,  Maturi M.,  2016, Weak gravitational lensing (\mn@eprint
  {arXiv} {1612.06535})

\bibitem[\protect\citeauthoryear{Bartelmann \& Schneider}{Bartelmann \&
  Schneider}{2001}]{bartelmann2001weak}
Bartelmann M.,  Schneider P.,  2001, Physics Reports, 340, 291

\bibitem[\protect\citeauthoryear{Baumann, Nicolis, Senatore  \&
  Zaldarriaga}{Baumann et~al.}{2012}]{PTBaumann_2012}
Baumann D.,  Nicolis A.,  Senatore L.,   Zaldarriaga M.,  2012, \mn@doi
  [Journal of Cosmology and Astroparticle Physics]
  {10.1088/1475-7516/2012/07/051}, 2012, 051

\bibitem[\protect\citeauthoryear{Bernardeau, Colombi, Gaztañaga  \&
  Scoccimarro}{Bernardeau et~al.}{2002}]{PTBERNARDEAU20021}
Bernardeau F.,  Colombi S.,  Gaztañaga E.,   Scoccimarro R.,  2002, \mn@doi
  [Physics Reports] {https://doi.org/10.1016/S0370-1573(02)00135-7}, 367, 1

\bibitem[\protect\citeauthoryear{Bernstein}{Bernstein}{2010}]{bernstein2010shape}
Bernstein G.~M.,  2010, Monthly Notices of the Royal Astronomical Society, 406,
  2793

\bibitem[\protect\citeauthoryear{Bernstein \& Jarvis}{Bernstein \&
  Jarvis}{2002}]{bernstein2002shapes}
Bernstein G.,  Jarvis M.,  2002, The Astronomical Journal, 123, 583

\bibitem[\protect\citeauthoryear{Beutler et~al.,}{Beutler
  et~al.}{2017}]{PTbeutler2017clustering}
Beutler F.,  et~al., 2017, Monthly Notices of the Royal Astronomical Society,
  464, 3409

\bibitem[\protect\citeauthoryear{Bird, Viel  \& Haehnelt}{Bird
  et~al.}{2012}]{10.1111/j.1365-2966.2011.20222.x}
Bird S.,  Viel M.,   Haehnelt M.~G.,  2012, \mn@doi [Monthly Notices of the
  Royal Astronomical Society] {10.1111/j.1365-2966.2011.20222.x}, 420, 2551

\bibitem[\protect\citeauthoryear{Blas, Garny  \& Konstandin}{Blas
  et~al.}{2014}]{PTBlas_2014}
Blas D.,  Garny M.,   Konstandin T.,  2014, \mn@doi [Journal of Cosmology and
  Astroparticle Physics] {10.1088/1475-7516/2014/01/010}, 2014, 010

\bibitem[\protect\citeauthoryear{Blas, Garny, Ivanov  \& Sibiryakov}{Blas
  et~al.}{2016}]{PTBlas_2016}
Blas D.,  Garny M.,  Ivanov M.~M.,   Sibiryakov S.,  2016, \mn@doi [Journal of
  Cosmology and Astroparticle Physics] {10.1088/1475-7516/2016/07/052}, 2016,
  052

\bibitem[\protect\citeauthoryear{Bridle \& King}{Bridle \&
  King}{2007}]{bridle2007dark}
Bridle S.,  King L.,  2007, New Journal of Physics, 9, 444

\bibitem[\protect\citeauthoryear{Cataneo, Lombriser, Heymans, Mead, Barreira,
  Bose  \& Li}{Cataneo et~al.}{2019}]{PTcataneo2019road}
Cataneo M.,  Lombriser L.,  Heymans C.,  Mead A.,  Barreira A.,  Bose S.,   Li
  B.,  2019, Monthly Notices of the Royal Astronomical Society, 488, 2121

\bibitem[\protect\citeauthoryear{Chevallier \& Polarski}{Chevallier \&
  Polarski}{2001}]{chevallier2001accelerating}
Chevallier M.,  Polarski D.,  2001, International Journal of Modern Physics D,
  10, 213

\bibitem[\protect\citeauthoryear{Choi et~al.,}{Choi
  et~al.}{2016}]{choi2016cfhtlens}
Choi A.,  et~al., 2016, Monthly Notices of the Royal Astronomical Society, 463,
  3737

\bibitem[\protect\citeauthoryear{Collaboration et~al.,}{Collaboration
  et~al.}{2020}]{euclidcollaboration2020euclid}
Collaboration E.,  et~al., 2020, Euclid preparation: IX. EuclidEmulator2 --
  Power spectrum emulation with massive neutrinos and self-consistent dark
  energy perturbations (\mn@eprint {arXiv} {2010.11288})

\bibitem[\protect\citeauthoryear{Cooray \& Sheth}{Cooray \&
  Sheth}{2002}]{COORAY20021}
Cooray A.,  Sheth R.,  2002, \mn@doi [Physics Reports]
  {https://doi.org/10.1016/S0370-1573(02)00276-4}, 372, 1

\bibitem[\protect\citeauthoryear{Crocce \& Scoccimarro}{Crocce \&
  Scoccimarro}{2006}]{PTPhysRevD.73.063519}
Crocce M.,  Scoccimarro R.,  2006, \mn@doi [Phys. Rev. D]
  {10.1103/PhysRevD.73.063519}, 73, 063519

\bibitem[\protect\citeauthoryear{Crocce, Scoccimarro  \& Bernardeau}{Crocce
  et~al.}{2012}]{PThttps://doi.org/10.1111/j.1365-2966.2012.22127.x}
Crocce M.,  Scoccimarro R.,   Bernardeau F.,  2012, \mn@doi [Monthly Notices of
  the Royal Astronomical Society]
  {https://doi.org/10.1111/j.1365-2966.2012.22127.x}, 427, 2537

\bibitem[\protect\citeauthoryear{{DES Collaboration} et~al.}{{DES
  Collaboration} et~al.}{2021}]{zuercher2021peaks}
{DES Collaboration} et~al., 2021, To be submitted to

\bibitem[\protect\citeauthoryear{Eisenstein \& Hu}{Eisenstein \&
  Hu}{1999}]{Eisenstein_1999}
Eisenstein D.~J.,  Hu W.,  1999, \mn@doi [The Astrophysical Journal]
  {10.1086/306640}, 511, 5

\bibitem[\protect\citeauthoryear{Fluri, Kacprzak, Lucchi, Refregier, Amara,
  Hofmann  \& Schneider}{Fluri et~al.}{2019}]{fluri2019cosmological}
Fluri J.,  Kacprzak T.,  Lucchi A.,  Refregier A.,  Amara A.,  Hofmann T.,
  Schneider A.,  2019, Physical Review D, 100, 063514

\bibitem[\protect\citeauthoryear{Foreman \& Senatore}{Foreman \&
  Senatore}{2016}]{PTforeman2016eft}
Foreman S.,  Senatore L.,  2016, Journal of Cosmology and Astroparticle
  Physics, 2016, 033

\bibitem[\protect\citeauthoryear{Foreman-Mackey, Hogg, Lang  \&
  Goodman}{Foreman-Mackey et~al.}{2013}]{Foreman_Mackey_2013}
Foreman-Mackey D.,  Hogg D.~W.,  Lang D.,   Goodman J.,  2013, \mn@doi
  [Publications of the Astronomical Society of the Pacific] {10.1086/670067},
  125, 306

\bibitem[\protect\citeauthoryear{Giannantonio, Porciani, Carron, Amara  \&
  Pillepich}{Giannantonio et~al.}{2012}]{giannantonio2012constraining}
Giannantonio T.,  Porciani C.,  Carron J.,  Amara A.,   Pillepich A.,  2012,
  Monthly Notices of the Royal Astronomical Society, 422, 2854

\bibitem[\protect\citeauthoryear{Giblin, Cataneo, Moews  \& Heymans}{Giblin
  et~al.}{2019}]{giblin2019road}
Giblin B.,  Cataneo M.,  Moews B.,   Heymans C.,  2019, Monthly Notices of the
  Royal Astronomical Society, 490, 4826

\bibitem[\protect\citeauthoryear{Gorski, Wandelt, Hansen, Hivon  \&
  Banday}{Gorski et~al.}{1999}]{gorski1999healpix}
Gorski K.~M.,  Wandelt B.~D.,  Hansen F.~K.,  Hivon E.,   Banday A.~J.,  1999,
  The HEALPix Primer (\mn@eprint {arXiv} {astro-ph/9905275})

\bibitem[\protect\citeauthoryear{Gorski, Hivon, Banday, Wandelt, Hansen,
  Reinecke  \& Bartelmann}{Gorski et~al.}{2005}]{gorski2005healpix}
Gorski K.~M.,  Hivon E.,  Banday A.~J.,  Wandelt B.~D.,  Hansen F.~K.,
  Reinecke M.,   Bartelmann M.,  2005, The Astrophysical Journal, 622, 759

\bibitem[\protect\citeauthoryear{{Hamilton}, {Kumar}, {Lu}  \&
  {Matthews}}{{Hamilton} et~al.}{1991}]{1991ApJ...374L...1H}
{Hamilton} A.~J.~S.,  {Kumar} P.,  {Lu} E.,   {Matthews} A.,  1991, \mn@doi
  [\apjl] {10.1086/186057}, \href
  {https://ui.adsabs.harvard.edu/abs/1991ApJ...374L...1H} {374, L1}

\bibitem[\protect\citeauthoryear{Hand, Feng, Beutler, Li, Modi, Seljak  \&
  Slepian}{Hand et~al.}{2018}]{Hand_2018}
Hand N.,  Feng Y.,  Beutler F.,  Li Y.,  Modi C.,  Seljak U.,   Slepian Z.,
  2018, \mn@doi [The Astronomical Journal] {10.3847/1538-3881/aadae0}, 156, 160

\bibitem[\protect\citeauthoryear{{Hartlap, J.}, {Simon, P.}  \& {Schneider,
  P.}}{{Hartlap, J.} et~al.}{2007}]{cov1}
{Hartlap, J.} {Simon, P.}  {Schneider, P.} 2007, \mn@doi [A\&A]
  {10.1051/0004-6361:20066170}, 464, 399

\bibitem[\protect\citeauthoryear{Heavens, Refregier  \& Heymans}{Heavens
  et~al.}{2000}]{heavens2000intrinsic}
Heavens A.,  Refregier A.,   Heymans C.,  2000, Monthly Notices of the Royal
  Astronomical Society, 319, 649

\bibitem[\protect\citeauthoryear{Heitmann, Higdon, White, Habib, Williams,
  Lawrence  \& Wagner}{Heitmann et~al.}{2009}]{heitmann2009coyote}
Heitmann K.,  Higdon D.,  White M.,  Habib S.,  Williams B.~J.,  Lawrence E.,
  Wagner C.,  2009, The Astrophysical Journal, 705, 156

\bibitem[\protect\citeauthoryear{Heitmann, White, Wagner, Habib  \&
  Higdon}{Heitmann et~al.}{2010}]{heitmann2010coyote}
Heitmann K.,  White M.,  Wagner C.,  Habib S.,   Higdon D.,  2010, The
  Astrophysical Journal, 715, 104

\bibitem[\protect\citeauthoryear{Heitmann, Lawrence, Kwan, Habib  \&
  Higdon}{Heitmann et~al.}{2013}]{heitmann2013coyote}
Heitmann K.,  Lawrence E.,  Kwan J.,  Habib S.,   Higdon D.,  2013, The
  Astrophysical Journal, 780, 111

\bibitem[\protect\citeauthoryear{Heitmann et~al.,}{Heitmann
  et~al.}{2016}]{Heitmann_2016}
Heitmann K.,  et~al., 2016, \mn@doi [The Astrophysical Journal]
  {10.3847/0004-637x/820/2/108}, 820, 108

\bibitem[\protect\citeauthoryear{Hildebrandt et~al.,}{Hildebrandt
  et~al.}{2020}]{hildebrandt2020kids+}
Hildebrandt H.,  et~al., 2020, Astronomy \& Astrophysics, 633, A69

\bibitem[\protect\citeauthoryear{Hinshaw et~al.,}{Hinshaw
  et~al.}{2013}]{Hinshaw_2013}
Hinshaw G.,  et~al., 2013, \mn@doi [The Astrophysical Journal Supplement
  Series] {10.1088/0067-0049/208/2/19}, 208, 19

\bibitem[\protect\citeauthoryear{Hirata \& Seljak}{Hirata \&
  Seljak}{2004}]{hirata2004intrinsic}
Hirata C.~M.,  Seljak U.,  2004, Physical Review D, 70, 063526

\bibitem[\protect\citeauthoryear{Hirata et~al.,}{Hirata
  et~al.}{2004}]{hirata2004galaxy}
Hirata C.~M.,  et~al., 2004, Monthly Notices of the Royal Astronomical Society,
  353, 529

\bibitem[\protect\citeauthoryear{Hunter}{Hunter}{2007}]{matplotlib}
Hunter J.~D.,  2007, \mn@doi [Computing in Science Engineering]
  {10.1109/MCSE.2007.55}, 9, 90

\bibitem[\protect\citeauthoryear{Huterer, Takada, Bernstein  \& Jain}{Huterer
  et~al.}{2006}]{huterer2006systematic}
Huterer D.,  Takada M.,  Bernstein G.,   Jain B.,  2006, Monthly Notices of the
  Royal Astronomical Society, 366, 101

\bibitem[\protect\citeauthoryear{Jing, Zhang, Lin, Gao  \& Springel}{Jing
  et~al.}{2006}]{jing2006influence}
Jing Y.,  Zhang P.,  Lin W.,  Gao L.,   Springel V.,  2006, The Astrophysical
  Journal, 640, L119

\bibitem[\protect\citeauthoryear{Joachimi, Mandelbaum, Abdalla  \&
  Bridle}{Joachimi et~al.}{2011}]{joachimi2011constraints}
Joachimi B.,  Mandelbaum R.,  Abdalla F.,   Bridle S.,  2011, Astronomy \&
  Astrophysics, 527, A26

\bibitem[\protect\citeauthoryear{{Kaiser}}{{Kaiser}}{1992}]{1992ApJ...388..272K}
{Kaiser} N.,  1992, \mn@doi [\apj] {10.1086/171151}, \href
  {https://ui.adsabs.harvard.edu/abs/1992ApJ...388..272K} {388, 272}

\bibitem[\protect\citeauthoryear{{Kaiser}}{{Kaiser}}{1998}]{1998ApJ...498...26K}
{Kaiser} N.,  1998, \mn@doi [\apj] {10.1086/305515}, \href
  {https://ui.adsabs.harvard.edu/abs/1998ApJ...498...26K} {498, 26}

\bibitem[\protect\citeauthoryear{Kaiser \& Squires}{Kaiser \&
  Squires}{1993}]{kaiser1993mapping}
Kaiser N.,  Squires G.,  1993, The Astrophysical Journal, 404, 441

\bibitem[\protect\citeauthoryear{Kilbinger et~al.,}{Kilbinger
  et~al.}{2017}]{kilbinger2017precision}
Kilbinger M.,  et~al., 2017, Monthly Notices of the Royal Astronomical Society,
  472, 2126

\bibitem[\protect\citeauthoryear{Kitching, Alsing, Heavens, Jimenez, McEwen  \&
  Verde}{Kitching et~al.}{2017}]{kitching2017limits}
Kitching T.~D.,  Alsing J.,  Heavens A.~F.,  Jimenez R.,  McEwen J.~D.,   Verde
  L.,  2017, Monthly Notices of the Royal Astronomical Society, 469, 2737

\bibitem[\protect\citeauthoryear{Knabenhans et~al.,}{Knabenhans
  et~al.}{2019}]{Euclid2019}
Knabenhans M.,  et~al., 2019, \mn@doi [Monthly Notices of the Royal
  Astronomical Society] {10.1093/mnras/stz197}, 484, 5509–5529

\bibitem[\protect\citeauthoryear{Knabenhans, Brinckmann, Stadel, Schneider  \&
  Teyssier}{Knabenhans et~al.}{2021}]{knabenhans2021parameter}
Knabenhans M.,  Brinckmann T.,  Stadel J.,  Schneider A.,   Teyssier R.,  2021,
  arXiv preprint arXiv:2110.01488

\bibitem[\protect\citeauthoryear{Kodwani, Alonso  \& Ferreira}{Kodwani
  et~al.}{2018}]{kodwani2018effect}
Kodwani D.,  Alonso D.,   Ferreira P.,  2018, arXiv preprint arXiv:1811.11584

\bibitem[\protect\citeauthoryear{Lawrence et~al.,}{Lawrence
  et~al.}{2017}]{lawrence2017mira}
Lawrence E.,  et~al., 2017, The Astrophysical Journal, 847, 50

\bibitem[\protect\citeauthoryear{Lesgourgues}{Lesgourgues}{2011}]{lesgourgues2011cosmic}
Lesgourgues J.,  2011, The Cosmic Linear Anisotropy Solving System (CLASS) III:
  Comparision with CAMB for LambdaCDM (\mn@eprint {arXiv} {1104.2934})

\bibitem[\protect\citeauthoryear{Lewis}{Lewis}{2019}]{lewis2019getdist}
Lewis A.,  2019, GetDist: a Python package for analysing Monte Carlo samples
  (\mn@eprint {arXiv} {1910.13970})

\bibitem[\protect\citeauthoryear{{Limber}}{{Limber}}{1953}]{1953ApJ...117..134L}
{Limber} D.~N.,  1953, \mn@doi [\apj] {10.1086/145672}, \href
  {https://ui.adsabs.harvard.edu/abs/1953ApJ...117..134L} {117, 134}

\bibitem[\protect\citeauthoryear{Linder}{Linder}{2003}]{PhysRevLett.90.091301}
Linder E.~V.,  2003, \mn@doi [Phys. Rev. Lett.]
  {10.1103/PhysRevLett.90.091301}, 90, 091301

\bibitem[\protect\citeauthoryear{LoVerde \& Afshordi}{LoVerde \&
  Afshordi}{2008}]{PhysRevD.78.123506}
LoVerde M.,  Afshordi N.,  2008, \mn@doi [Phys. Rev. D]
  {10.1103/PhysRevD.78.123506}, 78, 123506

\bibitem[\protect\citeauthoryear{{Ma} \& {Fry}}{{Ma} \&
  {Fry}}{2000}]{2000ApJ...543..503M}
{Ma} C.-P.,  {Fry} J.~N.,  2000, \mn@doi [\apj] {10.1086/317146}, \href
  {https://ui.adsabs.harvard.edu/abs/2000ApJ...543..503M} {543, 503}

\bibitem[\protect\citeauthoryear{Mancini, Piras, Alsing, Joachimi  \&
  Hobson}{Mancini et~al.}{2021}]{mancini2021cosmopower}
Mancini A.~S.,  Piras D.,  Alsing J.,  Joachimi B.,   Hobson M.~P.,  2021,
  Monthly Notices of the Royal Astronomical Society

\bibitem[\protect\citeauthoryear{Martinelli et~al.,}{Martinelli
  et~al.}{2021}]{2021Euclid}
Martinelli M.,  et~al., 2021, \mn@doi [Astronomy & Astrophysics]
  {10.1051/0004-6361/202039835}, 649, A100

\bibitem[\protect\citeauthoryear{Mead, Peacock, Heymans, Joudaki  \&
  Heavens}{Mead et~al.}{2015}]{10.1093/mnras/stv2036}
Mead A.~J.,  Peacock J.~A.,  Heymans C.,  Joudaki S.,   Heavens A.~F.,  2015,
  \mn@doi [Monthly Notices of the Royal Astronomical Society]
  {10.1093/mnras/stv2036}, 454, 1958

\bibitem[\protect\citeauthoryear{Mead, Heymans, Lombriser, Peacock, Steele  \&
  Winther}{Mead et~al.}{2016}]{10.1093/mnras/stw681}
Mead A.~J.,  Heymans C.,  Lombriser L.,  Peacock J.~A.,  Steele O.~I.,
  Winther H.~A.,  2016, \mn@doi [Monthly Notices of the Royal Astronomical
  Society] {10.1093/mnras/stw681}, 459, 1468

\bibitem[\protect\citeauthoryear{Melchior \& Viola}{Melchior \&
  Viola}{2012}]{melchior2012means}
Melchior P.,  Viola M.,  2012, Monthly Notices of the Royal Astronomical
  Society, 424, 2757

\bibitem[\protect\citeauthoryear{Mohammed \& Seljak}{Mohammed \&
  Seljak}{2014}]{10.1093/mnras/stu1972}
Mohammed I.,  Seljak U.,  2014, \mn@doi [Monthly Notices of the Royal
  Astronomical Society] {10.1093/mnras/stu1972}, 445, 3382

\bibitem[\protect\citeauthoryear{Moser, Lorenz, Schmitt, R{\'e}fr{\'e}gier,
  Fluri, Sgier, Tarsitano  \& Heisenberg}{Moser
  et~al.}{2022}]{moser2022symbolic}
Moser B.,  Lorenz C.,  Schmitt U.,  R{\'e}fr{\'e}gier A.,  Fluri J.,  Sgier R.,
   Tarsitano F.,   Heisenberg L.,  2022, Astronomy and Computing, p. 100603

\bibitem[\protect\citeauthoryear{Nishimichi, Bernardeau  \& Taruya}{Nishimichi
  et~al.}{2016}]{PTNISHIMICHI2016247}
Nishimichi T.,  Bernardeau F.,   Taruya A.,  2016, \mn@doi [Physics Letters B]
  {https://doi.org/10.1016/j.physletb.2016.09.035}, 762, 247

\bibitem[\protect\citeauthoryear{Peacock \& Smith}{Peacock \&
  Smith}{2000}]{10.1046/j.1365-8711.2000.03779.x}
Peacock J.~A.,  Smith R.~E.,  2000, \mn@doi [Monthly Notices of the Royal
  Astronomical Society] {10.1046/j.1365-8711.2000.03779.x}, 318, 1144

\bibitem[\protect\citeauthoryear{Peebles}{Peebles}{1980}]{peebles1980large}
Peebles P.,  1980, Large-Scale Structure of the Universe by Phillip James Edwin
  Peebles. Princeton University Press

\bibitem[\protect\citeauthoryear{Percival et~al.,}{Percival
  et~al.}{2014}]{cov2}
Percival W.~J.,  et~al., 2014, \mn@doi [Monthly Notices of the Royal
  Astronomical Society] {10.1093/mnras/stu112}, 439, 2531

\bibitem[\protect\citeauthoryear{Petri, Haiman  \& May}{Petri
  et~al.}{2017}]{petri2017validity}
Petri A.,  Haiman Z.,   May M.,  2017, Physical Review D, 95, 123503

\bibitem[\protect\citeauthoryear{Potter, Stadel  \& Teyssier}{Potter
  et~al.}{2017}]{potter2017pkdgrav3}
Potter D.,  Stadel J.,   Teyssier R.,  2017, Computational Astrophysics and
  Cosmology, 4, 2

\bibitem[\protect\citeauthoryear{{Press} \& {Schechter}}{{Press} \&
  {Schechter}}{1974}]{1974ApJ...187..425P}
{Press} W.~H.,  {Schechter} P.,  1974, \mn@doi [\apj] {10.1086/152650}, \href
  {https://ui.adsabs.harvard.edu/abs/1974ApJ...187..425P} {187, 425}

\bibitem[\protect\citeauthoryear{Refregier, Kacprzak, Amara, Bridle  \&
  Rowe}{Refregier et~al.}{2012}]{refregier2012noise}
Refregier A.,  Kacprzak T.,  Amara A.,  Bridle S.,   Rowe B.,  2012, Monthly
  Notices of the Royal Astronomical Society, 425, 1951

\bibitem[\protect\citeauthoryear{Refregier, Gamper, Amara  \&
  Heisenberg}{Refregier et~al.}{2017}]{refregier2017pycosmo}
Refregier A.,  Gamper L.,  Amara A.,   Heisenberg L.,  2017, PyCosmo: An
  Integrated Cosmological Boltzmann Solver (\mn@eprint {arXiv} {1708.05177})

\bibitem[\protect\citeauthoryear{Rudd, Zentner  \& Kravtsov}{Rudd
  et~al.}{2008}]{rudd2008effects}
Rudd D.~H.,  Zentner A.~R.,   Kravtsov A.~V.,  2008, The Astrophysical Journal,
  672, 19

\bibitem[\protect\citeauthoryear{Schneider et~al.,}{Schneider
  et~al.}{2016}]{Schneider_2016}
Schneider A.,  et~al., 2016, \mn@doi [Journal of Cosmology and Astroparticle
  Physics] {10.1088/1475-7516/2016/04/047}, 2016, 047

\bibitem[\protect\citeauthoryear{Scoccimarro, Sheth, Hui  \& Jain}{Scoccimarro
  et~al.}{2001}]{Scoccimarro_2001}
Scoccimarro R.,  Sheth R.~K.,  Hui L.,   Jain B.,  2001, \mn@doi [The
  Astrophysical Journal] {10.1086/318261}, 546, 20

\bibitem[\protect\citeauthoryear{Seljak}{Seljak}{2000}]{10.1046/j.1365-8711.2000.03715.x}
Seljak U.,  2000, \mn@doi [Monthly Notices of the Royal Astronomical Society]
  {10.1046/j.1365-8711.2000.03715.x}, 318, 203

\bibitem[\protect\citeauthoryear{Seljak \& Vlah}{Seljak \&
  Vlah}{2015}]{PhysRevD.91.123516}
Seljak U. c.~v.,  Vlah Z.,  2015, \mn@doi [Phys. Rev. D]
  {10.1103/PhysRevD.91.123516}, 91, 123516

\bibitem[\protect\citeauthoryear{Sgier, R{\'e}fr{\'e}gier, Amara  \&
  Nicola}{Sgier et~al.}{2019}]{sgier2019fast}
Sgier R.~J.,  R{\'e}fr{\'e}gier A.,  Amara A.,   Nicola A.,  2019, Journal of
  Cosmology and Astroparticle Physics, 2019, 044

\bibitem[\protect\citeauthoryear{{Sheth} \& {Tormen}}{{Sheth} \&
  {Tormen}}{1999}]{1999MNRAS.308..119S}
{Sheth} R.~K.,  {Tormen} G.,  1999, \mn@doi [\mnras]
  {10.1046/j.1365-8711.1999.02692.x}, \href
  {https://ui.adsabs.harvard.edu/abs/1999MNRAS.308..119S} {308, 119}

\bibitem[\protect\citeauthoryear{Smail, Hogg, Yan  \& Cohen}{Smail
  et~al.}{1995}]{smail1995deep}
Smail I.,  Hogg D.~W.,  Yan L.,   Cohen J.~G.,  1995, The Astrophysical
  Journal, 449, L105

\bibitem[\protect\citeauthoryear{Smith et~al.,}{Smith
  et~al.}{2003}]{10.1046/j.1365-8711.2003.06503.x}
Smith R.~E.,  et~al., 2003, \mn@doi [Monthly Notices of the Royal Astronomical
  Society] {10.1046/j.1365-8711.2003.06503.x}, 341, 1311

\bibitem[\protect\citeauthoryear{Springel}{Springel}{2005a}]{gadget-2}
Springel V.,  2005a, Monthly notices of the royal astronomical society, 364,
  1105

\bibitem[\protect\citeauthoryear{Springel}{Springel}{2005b}]{gedget2}
Springel V.,  2005b, \mn@doi [Monthly Notices of the Royal Astronomical
  Society] {10.1111/j.1365-2966.2005.09655.x}, 364, 1105

\bibitem[\protect\citeauthoryear{Springel, Yoshida  \& White}{Springel
  et~al.}{2001}]{gadget}
Springel V.,  Yoshida N.,   White S.~D.,  2001, New Astronomy, 6, 79

\bibitem[\protect\citeauthoryear{Springel, Pakmor, Zier  \& Reinecke}{Springel
  et~al.}{2020}]{springel2020simulating}
Springel V.,  Pakmor R.,  Zier O.,   Reinecke M.,  2020, Simulating cosmic
  structure formation with the GADGET-4 code (\mn@eprint {arXiv} {2010.03567})

\bibitem[\protect\citeauthoryear{Takahashi, Sato, Nishimichi, Taruya  \&
  Oguri}{Takahashi et~al.}{2012}]{Takahashi_2012}
Takahashi R.,  Sato M.,  Nishimichi T.,  Taruya A.,   Oguri M.,  2012, The
  Astrophysical Journal, 761, 152

\bibitem[\protect\citeauthoryear{Tarsitano et~al.,}{Tarsitano
  et~al.}{2020}]{tarsitano2020predicting}
Tarsitano F.,  et~al., 2020, Predicting Cosmological Observables with PyCosmo
  (\mn@eprint {arXiv} {2005.00543})

\bibitem[\protect\citeauthoryear{Tram, Brandbyge, Dakin  \& Hannestad}{Tram
  et~al.}{2019}]{tram2019fully}
Tram T.,  Brandbyge J.,  Dakin J.,   Hannestad S.,  2019, Journal of Cosmology
  and Astroparticle Physics, 2019, 022

\bibitem[\protect\citeauthoryear{Troxel et~al.,}{Troxel
  et~al.}{2018a}]{PhysRevD.98.043528}
Troxel M.~A.,  et~al., 2018a, \mn@doi [Phys. Rev. D]
  {10.1103/PhysRevD.98.043528}, 98, 043528

\bibitem[\protect\citeauthoryear{Troxel et~al.,}{Troxel
  et~al.}{2018b}]{troxel2018dark}
Troxel M.~A.,  et~al., 2018b, Physical Review D, 98, 043528

\bibitem[\protect\citeauthoryear{Virtanen et~al.,}{Virtanen
  et~al.}{2020}]{Virtanen2020Mar}
Virtanen P.,  et~al., 2020, \mn@doi [Nat. Methods] {10.1038/s41592-019-0686-2},
  17, 261

\bibitem[\protect\citeauthoryear{Wallis, McEwen, Kitching, Leistedt  \&
  Plouviez}{Wallis et~al.}{2017}]{wallis2017mapping}
Wallis C.~G.,  McEwen J.~D.,  Kitching T.~D.,  Leistedt B.,   Plouviez A.,
  2017, arXiv preprint arXiv:1703.09233

\bibitem[\protect\citeauthoryear{Zonca, Singer, Lenz, Reinecke, Rosset, Hivon
  \& Gorski}{Zonca et~al.}{2019}]{zonca2019healpy}
Zonca A.,  Singer L.~P.,  Lenz D.,  Reinecke M.,  Rosset C.,  Hivon E.,
  Gorski K.~M.,  2019, Journal of Open Source Software, 4, 1298

\bibitem[\protect\citeauthoryear{Z{\"u}rcher, Fluri, Sgier, Kacprzak  \&
  Refregier}{Z{\"u}rcher et~al.}{2021}]{zurcher2021cosmological}
Z{\"u}rcher D.,  Fluri J.,  Sgier R.,  Kacprzak T.,   Refregier A.,  2021,
  Journal of Cosmology and Astroparticle Physics, 2021, 028

\bibitem[\protect\citeauthoryear{d'Amico, Gleyzes, Kokron, Markovic, Senatore,
  Zhang, Beutler  \& Gil-Marín}{d'Amico et~al.}{2020}]{PTheory2020}
d'Amico G.,  Gleyzes J.,  Kokron N.,  Markovic K.,  Senatore L.,  Zhang P.,
  Beutler F.,   Gil-Marín H.,  2020, \mn@doi [Journal of Cosmology and
  Astroparticle Physics] {10.1088/1475-7516/2020/05/005}, 2020, 005–005

\bibitem[\protect\citeauthoryear{van~der Walt, Colbert  \& Varoquaux}{van~der
  Walt et~al.}{2011}]{numpy}
van~der Walt S.,  Colbert S.~C.,   Varoquaux G.,  2011, \mn@doi [Computing in
  Science Engineering] {10.1109/MCSE.2011.37}, 13, 22

\makeatother
\end{thebibliography}

%%%%%%%%%%%%%%%%% APPENDICES %%%%%%%%%%%%%%%%%%%%%

\appendix
\section{Power spectrum comparison}\label{sec:Appendix}
\begin{figure*}%[！htpb]/[H]
\centering
\includegraphics[width=18cm,height=12cm]{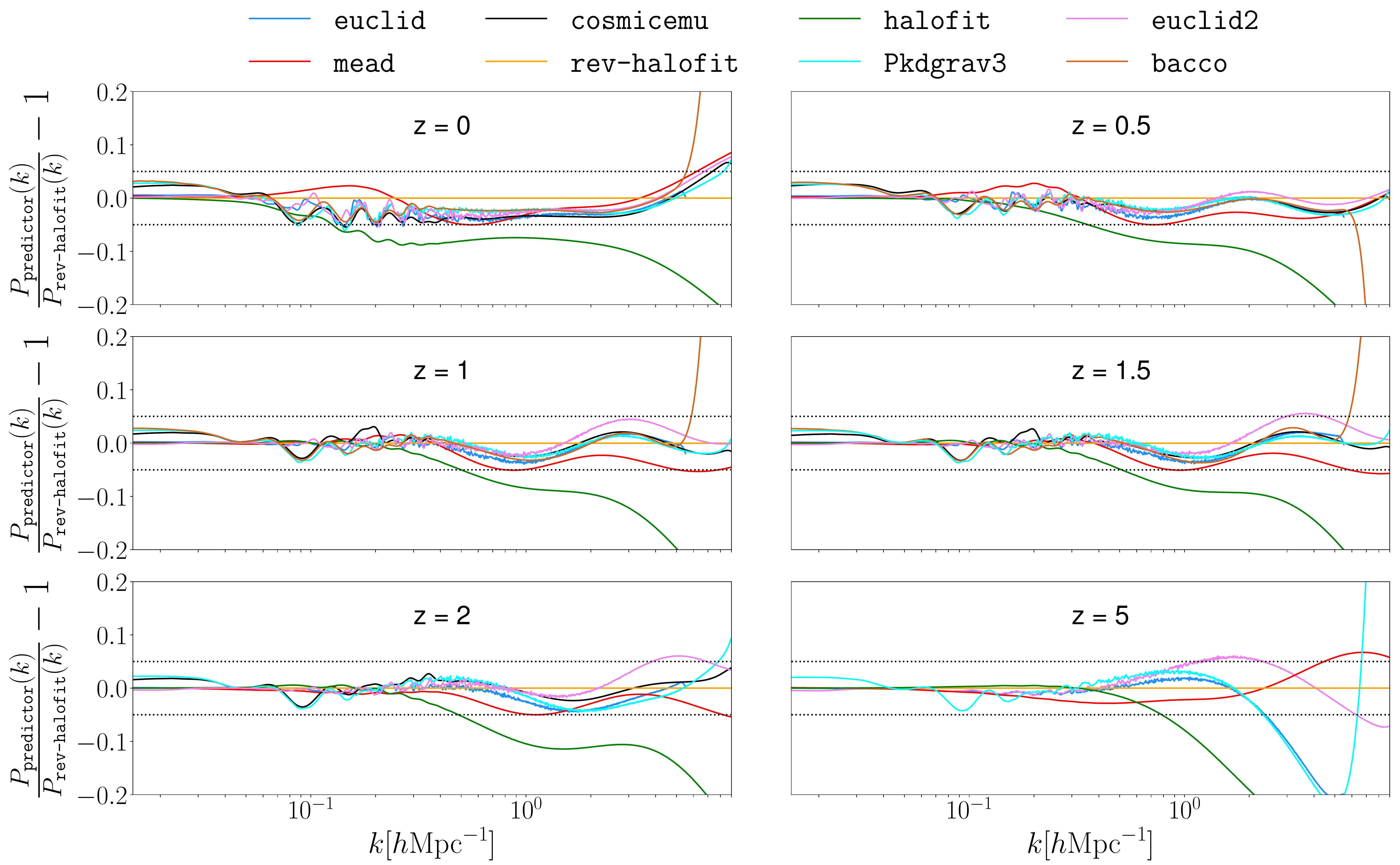}
\caption{The comparison of the dark-matter-only non linear $P(k)$ of different predictors at different redshifts ($z=0,0.5,1,1.5,2$ and $5$), subtracted and divided by $\texttt{rev-halofit}$ as reference. $\texttt{BaccoEmulator}$ and $\texttt{CosmicEmulator}$ are not valid for $z>3$, so we do not take them into comparison for $z=5$.}
\label{fig:power spectrum comparison2}
\end{figure*}

In this section we present the comparison of the non-linear power spectrum for all redshifts, as shown in Figure~\ref{fig:power spectrum comparison2}.
\\
\section{Cosmological Parameter Constraints}\label{sec:Appendix2}
The summary of constraints on $\{S_8,\Omega_{\rm{m}},n_{\rm{s}},h,w_0\}$ is concluded in this section, shown in Table~\ref{tab:S8 deviation all2}.

\begin{table*}%llllllllllllll
\begin{center}
\begin{tabular}{cccccccccccc}
\hline
Survey & Predictor&$S_8$&\scriptsize($\sigma$) &$\Omega_{\rm{m}}$&\scriptsize($\sigma$) &$n_{\rm{s}}$&\scriptsize($\sigma$) &$h$&\scriptsize($\sigma$) &$w_0$&\scriptsize($\sigma$) \\
Cosmology&\scriptsize ref: \texttt{rev-halofit} \\ 
\hline
\multirow{6}{*}{stage-\uppercase\expandafter{\romannumeral3}}&\scriptsize$\texttt{rev-halofit} $&\tiny$0.8147^{+0.0241}_{-0.0203}$&&\tiny$0.288^{+0.0817}_{-0.0662}$&&\tiny$0.9741^{+0.2489}_{-0.1475}$&&\tiny$0.6736^{+0.5642}_{-0.4172}$&&&\\
&\scriptsize$\texttt{mead} $&\tiny$0.8035^{+0.0269}_{-0.0202}$&\tiny$0.33$&\tiny$0.2996^{+0.0848}_{-0.0698}$&\tiny$0.11$&\tiny$0.9144^{+0.2425}_{-0.1562}$&\tiny$0.21$&\tiny$0.6859^{+0.5779}_{-0.4358}$&\tiny$0.02$&&\\
\multirow{4}{*}{$\Lambda \texttt{CDM}$}&\scriptsize$\texttt{halofit} $&\tiny$0.7946^{+0.0292}_{-0.0201}$&\tiny$0.57$&\tiny$0.2884^{+0.0783}_{-0.074}$&\tiny$0.0$&\tiny$0.9196^{+0.2566}_{-0.1662}$&\tiny$0.18$&\tiny$0.6777^{+0.6198}_{-0.4339}$&\tiny$0.01$&&\\
&\scriptsize$\texttt{euclid} $&\tiny$0.8083^{+0.0256}_{-0.0201}$&\tiny$0.2$&\tiny$0.2987^{+0.0831}_{-0.0709}$&\tiny$0.1$&\tiny$0.9547^{+0.2428}_{-0.1562}$&\tiny$0.07$&\tiny$0.6644^{+0.5612}_{-0.4159}$&\tiny$0.01$&&\\
&\scriptsize$\texttt{cosmicemu} $&\tiny$0.8047^{+0.0285}_{-0.018}$&\tiny$0.29$&\tiny$0.2916^{+0.0789}_{-0.0741}$&\tiny$0.03$&\tiny$0.9332^{+0.2613}_{-0.1399}$&\tiny$0.14$&\tiny$0.7452^{+0.5542}_{-0.486}$&\tiny$0.1$&&\\
&\scriptsize$\texttt{euclid2} $&\tiny$0.8031^{+0.0269}_{-0.0177}$&\tiny$0.34$&\tiny$0.2887^{+0.0835}_{-0.0679}$&\tiny$0.01$&\tiny$0.9184^{+0.2496}_{-0.1316}$&\tiny$0.19$&\tiny$0.7467^{+0.5503}_{-0.4853}$&\tiny$0.1$&&\\
\hline
\multirow{6}{*}{stage-\uppercase\expandafter{\romannumeral3}}&\scriptsize$\texttt{rev-halofit} $&\tiny$0.8165^{+0.0433}_{-0.0661}$&&\tiny$0.2846^{+0.092}_{-0.09}$&&\tiny$0.9164^{+0.5799}_{-0.3511}$&&\tiny$0.7868^{+0.9823}_{-0.5347}$&&\tiny$-0.9242^{+0.4704}_{-2.294}$&\\
&\scriptsize$\texttt{mead} $&\tiny$0.7947^{+0.0497}_{-0.0588}$&\tiny$0.26$&\tiny$0.31^{+0.0824}_{-0.1022}$&\tiny$0.18$&\tiny$0.9768^{+0.5012}_{-0.4514}$&\tiny$0.08$&\tiny$0.6527^{+1.0402}_{-0.4233}$&\tiny$0.11$&\tiny$-1.139^{+0.647}_{-2.2626}$&\tiny$0.09$\\
\multirow{4}{*}{$w \texttt{CDM}$}&\scriptsize$\texttt{halofit} $&\tiny$0.7879^{+0.0517}_{-0.0612}$&\tiny$0.34$&\tiny$0.2968^{+0.0787}_{-0.1011}$&\tiny$0.09$&\tiny$0.9919^{+0.4913}_{-0.4914}$&\tiny$0.1$&\tiny$0.6192^{+1.1863}_{-0.3779}$&\tiny$0.13$&\tiny$-1.1333^{+0.6581}_{-2.3122}$&\tiny$0.09$\\
&\scriptsize$\texttt{euclid} $&\tiny$0.7977^{+0.0545}_{-0.0542}$&\tiny$0.22$&\tiny$0.3049^{+0.085}_{-0.1017}$&\tiny$0.15$&\tiny$1.032^{+0.4723}_{-0.4813}$&\tiny$0.15$&\tiny$0.6209^{+1.1393}_{-0.3873}$&\tiny$0.13$&\tiny$-1.1886^{+0.7187}_{-2.1508}$&\tiny$0.11$\\
&\scriptsize$\texttt{cosmicemu} $&\tiny$0.7982^{+0.0504}_{-0.0572}$&\tiny$0.22$&\tiny$0.2931^{+0.0921}_{-0.0969}$&\tiny$0.06$&\tiny$1.0031^{+0.5093}_{-0.4918}$&\tiny$0.11$&\tiny$0.6688^{+1.2915}_{-0.4301}$&\tiny$0.08$&\tiny$-1.1408^{+0.6926}_{-2.3046}$&\tiny$0.09$\\
&\scriptsize$\texttt{euclid2} $&\tiny$0.8018^{+0.0461}_{-0.0627}$&\tiny$0.18$&\tiny$0.2896^{+0.0928}_{-0.0877}$&\tiny$0.04$&\tiny$0.9272^{+0.5729}_{-0.3921}$&\tiny$0.02$&\tiny$0.7461^{+1.0126}_{-0.5134}$&\tiny$0.04$&\tiny$-1.0254^{+0.5498}_{-2.2745}$&\tiny$0.04$\\
\hline
\multirow{6}{*}{stage-\uppercase\expandafter{\romannumeral4}}&\scriptsize$\texttt{rev-halofit} $&\tiny$0.8135^{+0.0023}_{-0.0024}$&&\tiny$0.2915^{+0.0077}_{-0.0084}$&&\tiny$0.9696^{+0.0178}_{-0.0192}$&&\tiny$0.6889^{+0.0481}_{-0.0433}$&&&\\
&\scriptsize$\texttt{mead} $&\tiny$0.8028^{+0.0027}_{-0.0026}$&\tiny$2.96$&\tiny$0.3008^{+0.0094}_{-0.0074}$&\tiny$0.87$&\tiny$0.9021^{+0.0193}_{-0.0189}$&\tiny$2.48$&\tiny$0.7181^{+0.0495}_{-0.0441}$&\tiny$0.45$&&\\
\multirow{4}{*}{$\Lambda \texttt{CDM}$}&\scriptsize$\texttt{halofit} $&\tiny$0.7944^{+0.002}_{-0.0029}$&\tiny$6.11$&\tiny$0.2856^{+0.0097}_{-0.0064}$&\tiny$0.46$&\tiny$0.9054^{+0.0203}_{-0.0197}$&\tiny$2.3$&\tiny$0.7134^{+0.0494}_{-0.05}$&\tiny$0.35$&&\\
&\scriptsize$\texttt{euclid} $&\tiny$0.8094^{+0.0018}_{-0.003}$&\tiny$1.37$&\tiny$0.2917^{+0.0079}_{-0.0084}$&\tiny$0.02$&\tiny$0.9497^{+0.0198}_{-0.0193}$&\tiny$0.72$&\tiny$0.7058^{+0.0505}_{-0.0479}$&\tiny$0.25$&&\\
&\scriptsize$\texttt{euclid2} $&\tiny$0.8058^{+0.0017}_{-0.0032}$&\tiny$2.62$&\tiny$0.2926^{+0.0079}_{-0.0084}$&\tiny$0.1$&\tiny$0.9402^{+0.0195}_{-0.0206}$&\tiny$1.07$&\tiny$0.6958^{+0.0519}_{-0.0475}$&\tiny$0.1$&&\\
\hline
\multirow{6}{*}{stage-\uppercase\expandafter{\romannumeral4}}&\scriptsize$\texttt{rev-halofit} $&\tiny$0.8127^{+0.0079}_{-0.0063}$&&\tiny$0.2909^{+0.0095}_{-0.0086}$&&\tiny$0.9741^{+0.045}_{-0.0555}$&&\tiny$0.6884^{+0.0599}_{-0.052}$&&\tiny$-1.0127^{+0.1171}_{-0.1046}$&\\
&\scriptsize$\texttt{mead} $&\tiny$0.7968^{+0.0067}_{-0.0069}$&\tiny$1.73$&\tiny$0.2979^{+0.0106}_{-0.0092}$&\tiny$0.53$&\tiny$0.9426^{+0.0431}_{-0.0466}$&\tiny$0.45$&\tiny$0.698^{+0.0563}_{-0.0449}$&\tiny$0.13$&\tiny$-1.106^{+0.1107}_{-0.1163}$&\tiny$0.61$\\
\multirow{4}{*}{$w \texttt{CDM}$}&\scriptsize$\texttt{halofit} $&\tiny$0.7902^{+0.007}_{-0.0073}$&\tiny$2.39$&\tiny$0.2856^{+0.0093}_{-0.0096}$&\tiny$0.42$&\tiny$0.9306^{+0.047}_{-0.0479}$&\tiny$0.6$&\tiny$0.6986^{+0.0606}_{-0.0507}$&\tiny$0.13$&\tiny$-1.0646^{+0.1069}_{-0.1197}$&\tiny$0.35$\\
&\scriptsize$\texttt{euclid} $&\tiny$0.8061^{+0.0073}_{-0.0072}$&\tiny$0.68$&\tiny$0.2908^{+0.0088}_{-0.0094}$&\tiny$0.01$&\tiny$0.9671^{+0.0574}_{-0.053}$&\tiny$0.09$&\tiny$0.6968^{+0.0609}_{-0.0604}$&\tiny$0.1$&\tiny$-1.046^{+0.1142}_{-0.1288}$&\tiny$0.22$\\
&\scriptsize$\texttt{euclid2} $&\tiny$0.7996^{+0.0078}_{-0.0069}$&\tiny$1.31$&\tiny$0.2901^{+0.0099}_{-0.0094}$&\tiny$0.06$&\tiny$0.9791^{+0.0515}_{-0.0588}$&\tiny$0.07$&\tiny$0.6711^{+0.0657}_{-0.0548}$&\tiny$0.21$&\tiny$-1.0965^{+0.1288}_{-0.1255}$&\tiny$0.51$\\
\hline

\end{tabular}
\caption{Complet numerical constraints on the cosmological parameters corresponding to the contours in Figure~\ref{fig:5params stage3}, \ref{fig:5params stage4}, \ref{fig:6params stage3}, and \ref{fig:6params stage4}. For each predictor, the $\sigma$s show the theoretical discrepancies for each parameter, compared to the reference one.}
\label{tab:S8 deviation all2}
\end{center}
\end{table*}
%%%%%%%%%%%%%%%%%%%%%%%%%%%%%%%%%%%%%%%%%%%%%%%%%%

% Don't change these lines
\bsp	% typesetting comment
\label{lastpage}
\end{document}